\documentclass[sn-mathphys,Numbered]{sn-jnl}

\theoremstyle{thmstyleone}%
\theoremstyle{thmstyletwo}%

\theoremstyle{thmstylethree}%

\usepackage{verbatim}
\usepackage{makecell}

\usepackage{lineno}

\begin{document}

\title[ ]{\textbf{Fast and reconfigurable sort-in-memory system enabled by memristors}}


\author[1]{\fnm{Lianfeng} \sur{Yu}}
\author*[1,2]{\fnm{Yaoyu} \sur{Tao}}\email{taoyaoyutyy@pku.edu.cn}
\author[1]{\fnm{Teng} \sur{Zhang}}
\author[1,3]{\fnm{Zeyu} \sur{Wang}}
\author[1,3]{\fnm{Xile} \sur{Wang}}
\author[1]{\fnm{Zelun} \sur{Pan}}
\author[1]{\fnm{Bowen} \sur{Wang}}
\author[1]{\fnm{Zhaokun} \sur{Jing}}
\author[1]{\fnm{Jiaxin} \sur{Liu}}
\author[1]{\fnm{Yuqi} \sur{Li}}
\author[1]{\fnm{Yihang} \sur{Zhu}}
\author[1,2]{\fnm{Bonan} \sur{Yan}}
\author*[1,2,3,4]{\fnm{Yuchao} \sur{Yang}}\email{yuchaoyang@pku.edu.cn}

\affil[1]{\orgdiv{ Beijing Advanced Innovation Center for Integrated Circuits, School of Integrated Circuits}, \orgname{Peking University}, \orgaddress{\city{Beijing}, \postcode{100871}, \country{China}}}
\affil[2]{\orgdiv{ Center for Brain Inspired Chips, Institute of Artificial Intelligence}, \orgname{Peking University}, \orgaddress{\city{Beijing}, \postcode{100871}, \country{China}}}
\affil[3]{\orgdiv{ Center for Brain Inspired Intelligence}, \orgname{Chinese Institute for Brain Research (CIBR)}, \orgaddress{\city{Beijing}, \postcode{102206}, \country{China}}}
\affil[4]{\orgdiv{ School of Electronic and Computer Engineering}, \orgname{Peking University}, \orgaddress{\city{Shenzhen}, \postcode{518055}, \country{China}}}

\maketitle


\renewcommand{\baselinestretch}{2.0}

\newpage
\noindent
{\fontsize{14pt}{14pt}\selectfont \textbf{Abstract}}
\vspace{6pt}

\noindent Sorting is fundamental and ubiquitous in modern computing systems. Hardware sorting systems are built based on comparison operations with Von Neumann architecture, but their performance are limited by the bandwidth between memory and comparison units and the performance of complementary metal–oxide–semiconductor (CMOS) based circuitry. Sort-in-memory (SIM) based on emerging memristors is desired but not yet available due to comparison operations that are challenging to be implemented within memristive memory. Here we report fast and reconfigurable SIM system enabled by digit read (DR) on 1-transistor-1-resistor (1T1R) memristor arrays. We develop DR tree node skipping (TNS) that support variable data quantity and data types, and extend TNS with multi-bank, bit-slice and multi-level strategies to enable cross-array TNS (CA-TNS) for practical adoptions. Experimented on benchmark sorting datasets, our memristor-enabled SIM system presents up to $3.32\times \sim 7.70 \times$ speedup, $6.23 \times \sim 183.5 \times$ energy efficiency improvement and $2.23 \times \sim 7.43 \times$ area reduction compared with state-of-the-art sorting systems. We apply such SIM system for shortest path search with Dijkstra's algorithm and neural network inference with in-situ pruning, demonstrating the capability in solving practical sorting tasks and the compatibility in integrating with other compute-in-memory (CIM) schemes. The comparison-free TNS/CA-TNS SIM enabled by memristors pushes sorting into a new paradigm of sort-in-memory for next-generation sorting systems. 

\newpage
\noindent{\fontsize{14pt}{14pt}\selectfont \textbf{Keywords}}
\vspace{6pt}

\noindent Sort-in-memory, memristor, comparison-free, digit read, tree node skipping, cross-array, in-situ pruning



\newpage
\section{Introduction}\label{sec1}

\begin{figure}[hbt!]
    \centering
    \includegraphics[width = 1\linewidth]{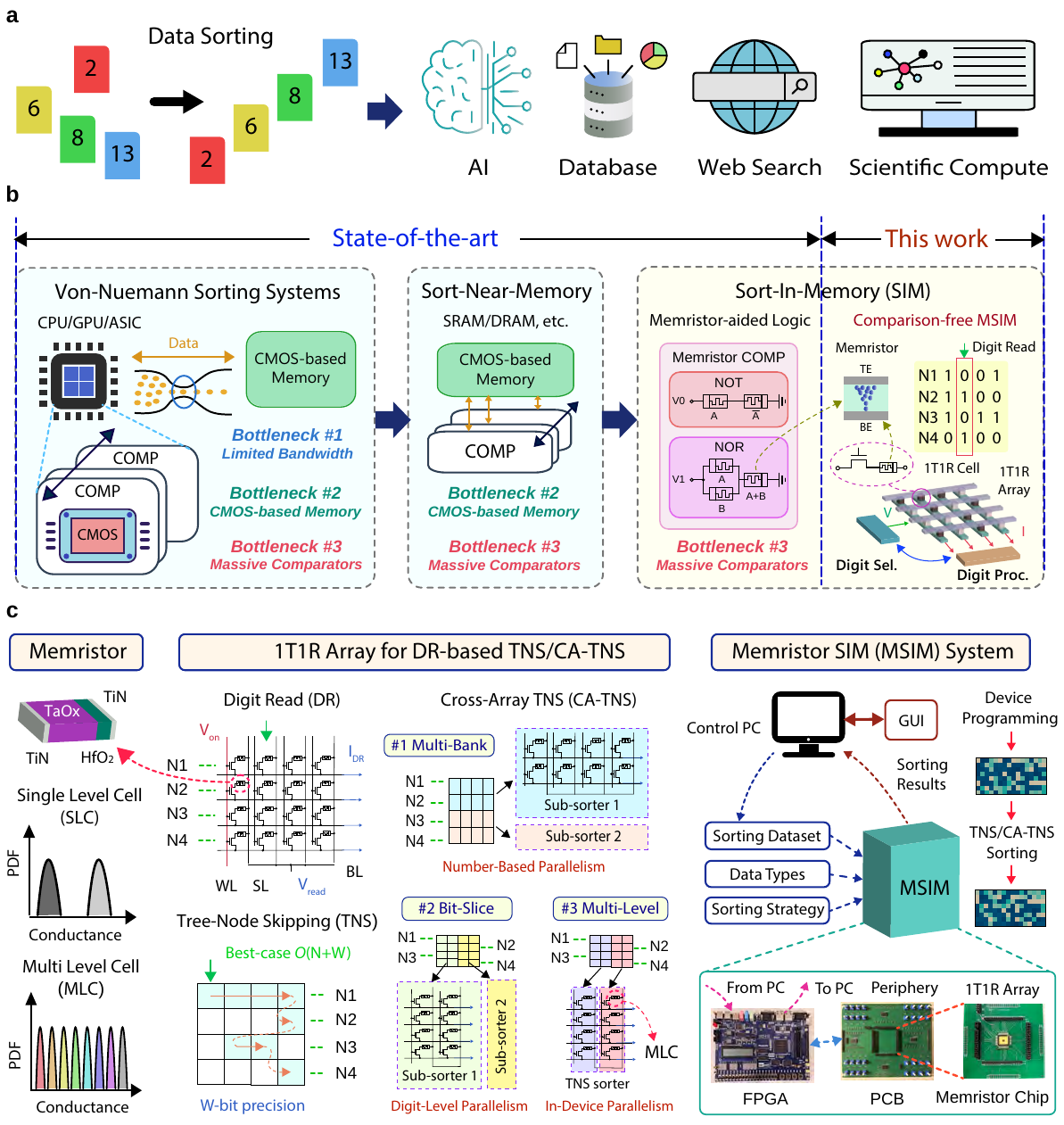}
    \caption{\textbf{Overview of sorting systems.} \textbf{a,} Sorting tasks are ubiquitous in numerous applications. \textbf{b,} CPU/GPU or ASIC-based sorting systems employ massive comparison units. The performance are limited by CMOS devices and the bandwidth between memory and comparison units. Sort-near-memory alleviates the bandwidth bottleneck. Sort-in-memory based on memristor-aided logic\cite{alam2022sorting} utilizes memristors but still relies on comparison operations, suffering from poor scalability, endurance and storage density. Our comparison-free MSIM with TNS/CA-TNS strategies resolves the three bottlenecks. \textbf{c,} Enabled by memristor devices, this article develops 1T1R array for digit read (DR), TNS/CA-TNS techniques for comparison-free SIM and an end-to-end MSIM system for practical demonstration.}
    \label{fig:introduction}
\end{figure}

Sorting is known to be a major performance bottleneck in numerous applications, including artificial intelligence \cite{raihan2020sparse,elsken2019neural,graves2016hybrid,petersen2021differentiable,kim2023neural,tao2021hima}, database \cite{taniar2002parallel, graefe2006implementing, govindaraju2006gputerasort, salamat2021nascent}, web search \cite{brin1998anatomy, guan2007eye} and scientific computing \cite{mankowitz2023faster,yang2021sorting,iram2023molecular,morris2013chemically,tkachenko2014optofluidic,arnold2006sorting}, as shown in \figurename~\ref{fig:introduction}a. It is used billions to trillions of times on any given day in modern computing systems \cite{mankowitz2023faster}. In fact, the most powerful supercomputers are sometimes used to perform large-scale sorting tasks in scenarios such as weather forecast \cite{george2014weather} or drug development \cite{shabaz2019sa}. To improve sorting performance, hardware sorting systems are designed mainly based on CPUs/GPUs or ASICs implementing comparison-based sorting algorithms such as merge sort \cite{cole1988parallel} or quick sort \cite{hoare1962quicksort}. Whereas remarkable progress has been achieved in the past, making further improvements in sorting performance is becoming increasingly challenging. This is because existing sorting systems usually rely on Von Neumann architecture with separate comparison units and memory, and utilize complex complementary metal–oxide–semiconductor (CMOS) based circuitry to execute comparison operations and store datasets and comparison results. The performance of these sorting systems can easily get saturated or degraded even with higher parallelism of comparison units.  

Recent advances in compute-in-memory (CIM) techniques allow computations to be completed within memory, minimizing data movements between memory and processing units \cite{wan2022compute, sebastian2020memory, yu2021compute, yan20221,wu202322nm,choi2023333tops,hu2022512gb}. Instead of using classical CMOS devices, emerging memristors offer significantly improved performance to carry out in-memory multiplications or accumulations \cite{hung20228, huang2023nonvolatile}, supporting tasks like artificial neural networks \cite{cai2020power, wang2022implementing,xue2021cmos,hung2021four,chiu2023cmos}, linear equation solvers \cite{khalid2019review} or differential equation solvers \cite{zidan2018general}. However, sort-in-memory (SIM) is considerably more challenging because they require highly-parallel comparison and select operations that are nontrivial to be implemented within memristive memory. Prior work propose memristor-aided logic \cite{alam2022sorting} that builds compare and select network within memristor arrays using memristor-based logic gates (NOT, NOR, etc.). However, large number of memristors are used to implement logic with frequent write operations, resulting in low storage density and short device lifetime (\figurename~\ref{fig:introduction}b). Moreover, as data quantity or data precision grows, the routing efforts increase significantly to assemble compare-and-swap (CAS) network, resulting in a poor scalability. To summarize, state-of-the-art memristor-based SIM still relies on comparison operations; hence, realizing comparison-free SIM using memristors is greatly desired to fundamentally improve sorting performance.

In this article, we report fast and reconfigurable memristor-based SIM (MSIM) system enabled by digit read (DR) on 1-transistor-1-resistor (1T1R) memristor arrays that can completely eliminate comparison operations and their relevant data movements. Min/max values are located iteratively by traversing the DR tree from most significant bit (MSB) to least significant bit (LSB). To reduce the number of DRs, we propose tree node skipping (TNS) that can dynamically records DR tree nodes to enable short-range traversal starting from an intermediate tree node (other than MSB) and ending at another intermediate tree node (before reaching LSB). Our TNS supports variable data precision (fixed-point or floating-point for any bits) and data types (unsigned, sign-and-magnitude and two's complement). To enhance TNS parallelism for practical adoptions, we further develop three cross-array TNS (CA-TNS) strategies: multi-bank strategy for number-based parallelism, bit-slice strategy for digit-level parallelism and multi-level strategy for in-device parallelism. We experimentally sort five benchmark sorting datasets of length 1024 and 32-bit numbers (random, normal, clustered, Kruskal's and MapReduce) using our MSIM system, and results demonstrate up to 6.91$\times$ speedup, 183.5$\times$ energy efficiency enhancement, and 4.76$\times$ area reduction over ASIC-based sorting systems using more advanced process technology. Applying such MSIM system for two representative real-world applications, shortest path search with Dijkstra's algorithm \cite{dijkstra2022note} and PointNet++ \cite{qi2017pointnet++} inference with run-time tunable sparsity, we demonstrate the capability of solving real-world sorting problems and the compatibility of integrating with other CIM techniques. The comparison-free SIM techniques enabled by memristors and the TNS/CA-TNS strategies result in a highly efficient and reconfigurable MSIM system, pushing sorting into a new era of sort-in-memory and demonstrating promising prospect for next-generation sorting system design.

\section{Memristor-Enabled Comparison-free SIM}\label{sec2} 

Our comparison-free MSIM system aims to effectively resolve three major bottlenecks (limited bandwidth, CMOS-based memory circuitry, and massive comparison units) in state-of-the-art sorting systems. It supports reconfiguration for different operating modes and different data types. \figurename~\ref{fig:introduction}c summarizes the innovations of this work in memristor device, 1T1R array design for DR-based TNS/CA-TNS and end-to-end MSIM system design.

\subsection{1T1R Memristor Array for Digit Read}

\begin{figure}[hbt!]
    \centering
    \includegraphics[width = 1\linewidth]{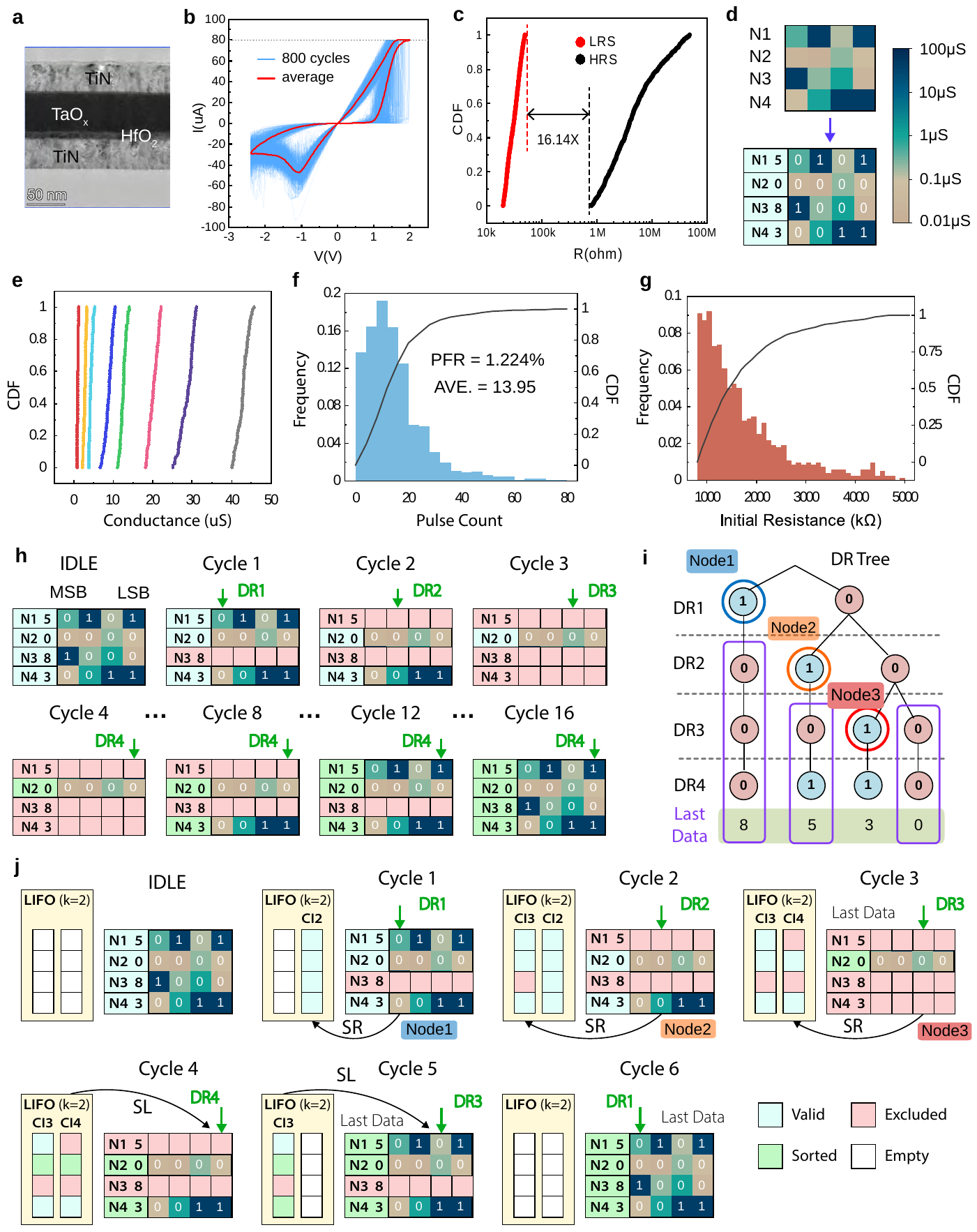}
    \caption{\textbf{Memristor programming and tree node skipping (TNS).} \textbf{a,} TEM image of memristor device. \textbf{b-g,} Experimental results of single-level-cell (SLC) and multi-level-cell (MLC) memristor programming. \textbf{h-j,} BTS, DR tree and TNS sorting for an example dataset of four 4-bit unsigned fixed-point numbers.}
    \label{fig:architecture}
\end{figure}

We first design and fabricate 1-transistor-1-resistor (1T1R) memristor array chip to enable comparison-free SIM techniques based on digit read (DR). The 1T1R array structure is described in \ref{fabricationsec11} and is taped out based on standard 180nm CMOS technology. \figurename~\ref{fig:architecture}a presents the transmission electron microscopes (TEM) image of our fabricated memristor device. We test the I-V performance (\figurename~\ref{fig:architecture}b) of our memristors under direct current (DC) scanning. With a set voltage of 2V and a reset voltage of 2.4V, our memristors show good switching characteristics with lowest ON/OFF switching ratio reaching 16.14$\times$ (\figurename~\ref{fig:architecture}c). Using DC scanning, we program an example dataset onto our 1T1R memristor array (\figurename~\ref{fig:architecture}d) for illustration of comparison-free SIM using DRs later. We further design a write-verify scheme (\figurename~\ref{fig:write_verify}a) to enable fast and accurate programming to the target conductance states to enable multi-level cells. Considering conductance overlap and programming efforts, we choose 8 conductance states (\figurename~\ref{fig:architecture}e) based on non-linear target conductance. \figurename~\ref{fig:architecture}f and \figurename~\ref{fig:architecture}g demonstrate the average programming efforts (in pulse count), programming failure rate (PFR) and the cumulative density function (CDF) when starting from different initial conductance.

We use the example in \figurename~\ref{fig:architecture}d for illustration of comparison-free SIM using DRs. Basic DR-based sorting (bit traversal SIM, or BTS\cite{prasad2021memristive}) traverses data numbers from their MSB to LSB to locate min/max values iteratively. \figurename~\ref{fig:architecture}h shows an example of sorting four unsigned 4-bit fixed-point numbers that takes 16 DRs to complete. BTS takes $O(N\times W)$ latency to sort a length-$N$ dataset of $W$-bit data precision. In DR-based SIM, DR of the $i$-th digit ($i = 1\rightarrow W$) is executed by applying read voltages on the memristor array followed by sense amplifiers. Suppose sorting of unsigned numbers in ascending order. The min search starts from the MSB: if bit '1' is encountered during DRs, the corresponding number is excluded (unless all numbers corresponding to that DR have bit '1'). The min search ends when the LSB is reached and the survival numbers correspond to the min values. Similarly, numbers that have bit '0' in DRs (unless all numbers corresponding to that DR have bit '0') can be successively excluded to locate max values. Using DR-based min/max search eliminates frequent write operations and improves data storage density compared to existing SIM techniques\cite{alam2022sorting}. The MSB-LSB traversal process can be regarded as tree node traversal of depth-$W$ DR tree as shown in \figurename~\ref{fig:architecture}i (Supplementary Section~\ref{Details_BTS} and \ref{Details_TNS}). A periphery is designed to support number exclusion (NE), DR and their associated logic (Supplementary Section~\ref{TNS_Hardware}). 

\subsection{Tree Node Skipping on Single Memristor Array}\label{subsec1}

\subsubsection{TNS Operation Flow and Hardware Design}\label{subsubsec2.2.1}

We observe that BTS introduces a large number of redundant DRs which are repeatedly executed. Unnecessary DRs can happen in three scenarios as follows: 1) Some DRs may have been processed previously for number exclusions, i.e., we do not need to exclude any new numbers for those digits in a min/max search iteration. For example, elements in pre-sort dataset may include many leading 0's where DRs on these leading 0's may be skipped; 2) DRs for repeating numbers in the dataset need redundant min/max search iterations that may be skipped for speedup; 3) DRs for the last number in the dataset can also be skipped. 

\begin{figure}[hbt!]
    \centering
    \includegraphics[width = 1\linewidth]{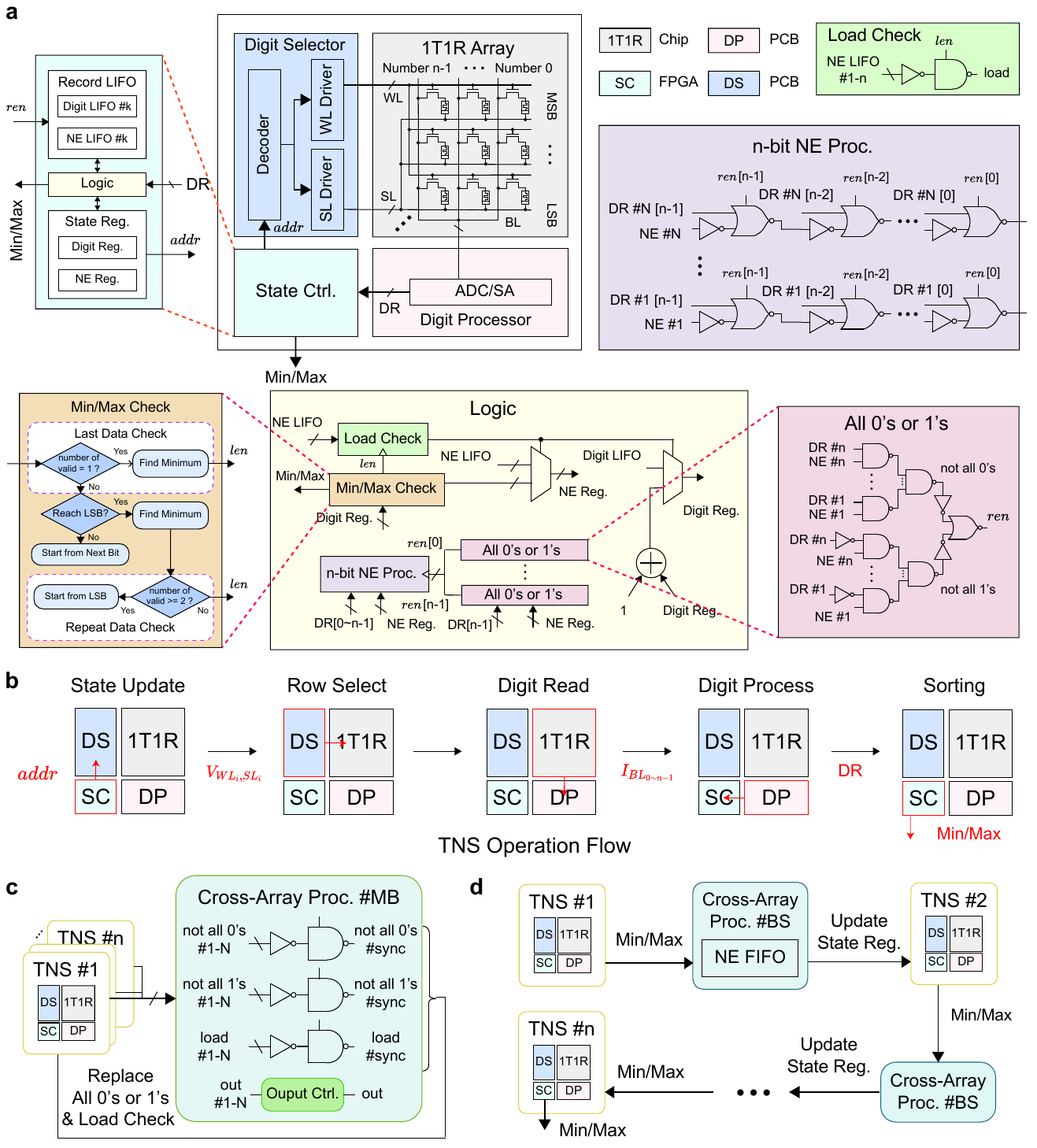}
    \caption{\textbf{TNS/CA-TNS Hardware architecture and operation flow.} \textbf{a,} Hardware design to support TNS and CA-TNS. \textbf{b,} TNS operation flow. \textbf{c-d,} Cross-array processor configured for multi-bank strategy and bit-slice strategy.}
    \label{fig:Fig3}
\end{figure}

In order to detect and skip redundant DRs in scenario 1, we propose to record the $k$ most recent tree nodes and their corresponding digit indexes. A straightforward implementation is to divide the process into two parts, one for tree node recording and the other for tree node reloading \cite{yu2022fast}. However, this implementation can only skip limited number of DRs mainly due to the separation of recording and loading operations. A new round of state recording can take place only after all currently stored tree nodes have been reloaded. Therefore, when reloading old tree nodes, we cannot record new tree nodes concurrently that can skip more DRs in a min/max search. To solve this problem, we develop tree node skipping (TNS) where recording and reloading can be executed simultaneously, ensuring that the recorded tree nodes are the ones that can skip the most DRs (the nodes circled in \figurename~\ref{fig:architecture}i and \figurename~\ref{fig:architecture}j). We also develop handling mechanisms for scenarios 2 and 3 to skip DRs for repeated numbers and last numbers in the dataset (Supplementary Section~\ref{Details_TNS}). \figurename~\ref{fig:architecture}j shows that sorting the same dataset as in \figurename~\ref{fig:architecture}h only takes 6 DRs using TNS instead of 16 DRs using BTS. \figurename~\ref{fig:Fig3}b summarizes the comparison-free TNS flow within a min/max search iteration. Pre-sort dataset are stored in a 1-transistor-1-resistor (1T1R) memristor array in binary form, where one dimension represents data numbers and the other dimension represents digit positions. DR starts from MSB for $1$st min/max search iteration. The detailed TNS process is described in Supplementary Section~\ref{Details_TNS}. It enables DR tree traversal starting from an intermediate tree node (other than MSB) and ending at another intermediate tree node (before reaching LSB), reducing the SIM latency from $O(N\times W)$ to $O(N+W)$ in the best case.   

To implement TNS, we design hardware consisting of four modules: 1T1R array chip, state controller (SC), digit selector (DS) and digit processor (DP) as shown in \figurename~\ref{fig:Fig3}a. Firstly, the state controller sends an address to digit selector, after which the decoder generates WL and SL signals to the 1T1R array. According to the device conductance and associated multi-level choices, sense amplifiers (SA) or analog-to-digital converters (ADC) are employed in digit processor to derive DR results. Secondly, DR results go to state controller that implements TNS logic (Supplementary Section~\ref{TNS_Hardware}). Finally, state controller determines the next digit to read or outputs the min/max value if found. The state controller is mainly built by three sub-blocks: state registers that stores digit indexes (digit reg.) and number status (NE reg.), length-$k$ last-in-first-out (LIFO) module that stores $k$ most recent tree nodes, and logic module that supports number exclusion, state recording and state reloading. The logic module also receives DR results and outputs the min/max of a search iteration. One may notice that the performance of TNS is closely related to the newly introduced parameter $k$. As $k$ increases, more tree nodes on DR tree are recorded and more DRs are likely to be skipped for better sorting speed, but the area and power consumption of LIFO and logic modules also grow. We study the detailed design trade-off between speed, area and energy efficiency for various $k$ in Supplementary Section~\ref{Details_Performance}.

\subsubsection{Periphery Design for Variable Data Types}\label{subsubsec2.2.2}

Our SIM system is also recongigurable to support variable data types in the literature including unsigned fixed-point numbers, two's complement fixed-point numbers or floating-point numbers and so on to meet the needs of real-world sorting applications. For unsigned fixed-point numbers, TNS iteratively traverses digits from upper digits to lower digits and excludes numbers with digit 1's or 0's to locate the min or max values, respectively. However, for two's complement fixed-point numbers, we need to extend our NE mechanisms to accommodate the additional sign bit. Taking sorting in ascending order for illustration, TNS needs to first exclude numbers with digit 0's at the sign bit, after which exclude numbers with digit 1's at successive digits. This is because the sign bit not only indicates whether a number is positive or negative, but also contributes to its magnitude as MSB. On the other hand, for floating-point representations, the sign bit only affects whether a number is positive or negative. Hence, for negative numbers (i.e., sign bit is 1), we need to exclude numbers with digit 0's at successive digits to locate min values; while for positive numbers (i.e., sign bit is 0), we need to exclude numbers with digit 1's at successive digits to locate min values. In addition to the data types mentioned above, other data types such as sign-and-magnitude can also be realized by extending NE logic as shown in Supplementary Section~\ref{TNS_Data_Types}. 

\subsection{Extending TNS on Multiple Memristor Arrays}\label{subsec2}

\begin{figure}[hbt!]
    \centering
    \includegraphics[width = 1\linewidth]{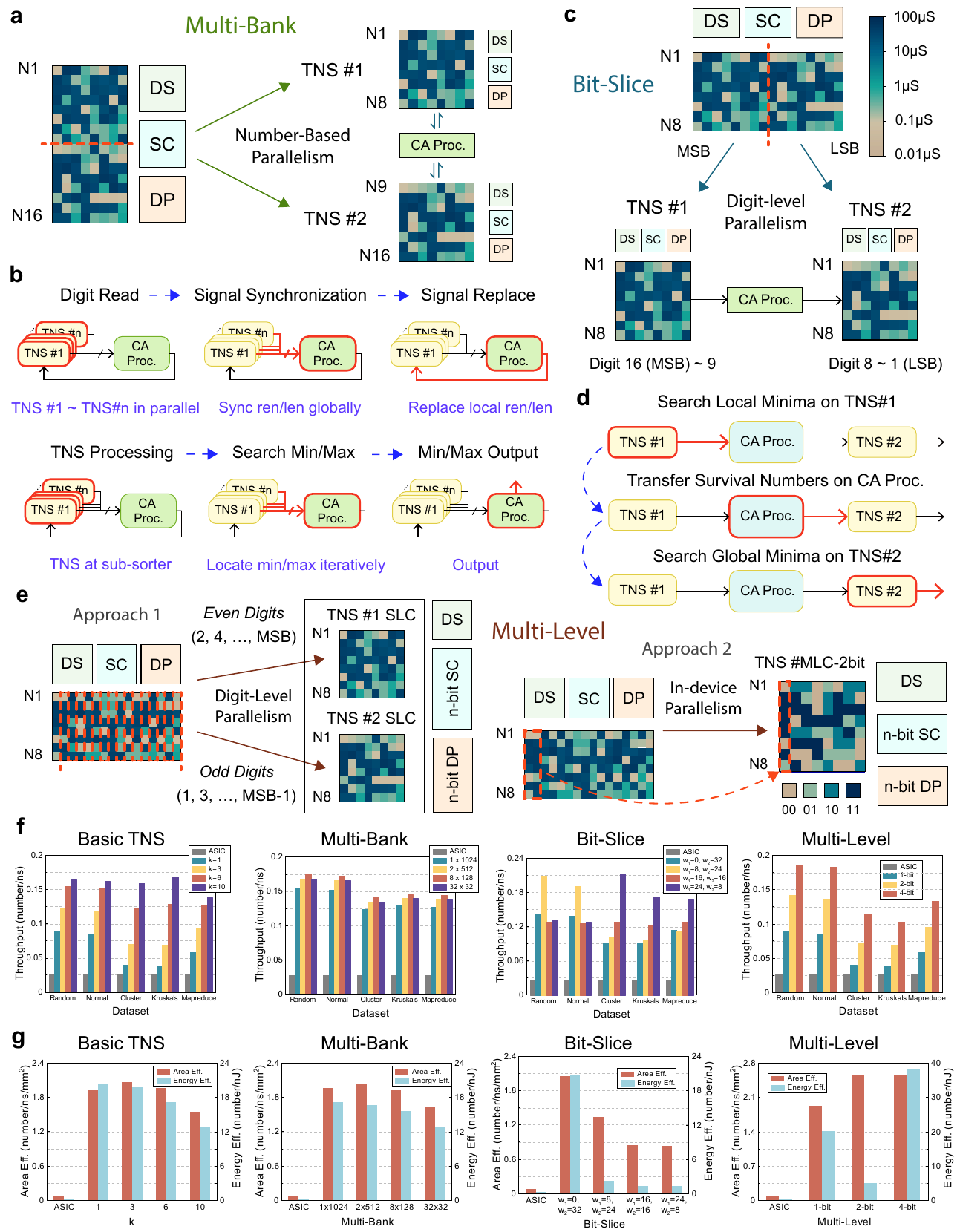}
    \caption{\textbf{Cross-array TNS (CA-TNS) strategies.} \textbf{a-e,} Multi-bank (MB), bit-slice (BS), and multi-level (ML) strategies. \textbf{f,} Sorting speed (number/$ns$), \textbf{g,} area and power of ASIC-based merge sorter (running at $1$GHz), basic TNS and CA-TNS (running at 312MHz $\sim$ 435MHz) on five benchmark sorting datasets. For illustration, we choose $k=6$ for MB, $k=4$ for BS and $k=1$ for ML strategies, respectively (Detailed study in Section~\ref{Details_Performance}).}
    \label{fig:ca_strategies}
\end{figure}

In real-world applications, pre-sort datasets are usually distributed and stored in multiple memory banks. To improve scalability and parallelism of MSIM techniques, we extend TNS across multiple memristor arrays and develop three cross-array TNS strategies as shown in \figurename~\ref{fig:ca_strategies}a-e. Our MSIM system can be configured for different strategies to meet different sorting requirements. A demonstration video (Supplementary Video 1) is provided for CA-TNS sorting.

\subsubsection{Multi-Bank Strategy}\label{subsubsec1}

The periphery described in \figurename~\ref{fig:Fig3} can theoretically support variable data quantity $N$ for sort-in-memory operations. However, in practical memristor array design, the parasitic resistance and capacitance connected to output bitline grow with 1T1R array size, causing increasing nonidealities. Hence, the size of practical 1T1R array is limited by manufacturing capability for tolerable nonidealities. On the other hand, complexity of TNS periphery increases super-linearly with $N$; therefore, it is unlikely to practically adopt our TNS for large-scale data sorting with big $N$. To overcome these challenges, we design a multi-bank (MB) strategy that supports sorting of practical dataset distributed in $n$ different memristor banks as shown in \figurename~\ref{fig:ca_strategies}a. Each memristor bank has its own periphery and can run as an independent sub-sorter of length $n_{mbi}$, where $n_{mbi}$ follows \eqref{mb_eq}:

\begin{equation}
\label{mb_eq}
\sum_{i = 1}^{n}n_{mbi} = N    
\end{equation}
 
An additional cross-array processor configured as multi-bank (MB) mode is developed to synchronize sorting of connected $n$ parallel sub-sorters, ensuring that all sub-sorters working simultaneously as a length-$N$ sorter (Supplementary Section~\ref{MB_Strategy}). Multi-bank cross-array strategy applied for length-$N$ dataset of $W$-bit numbers maintains a latency (in terms of number of DRs) equal to basic TNS as shown in \eqref{mb_sub_latency}, 

\begin{equation}
\label{mb_sub_latency}
T^{N,W}_{mb} = T^{N,W}_{\text{TNS}}
\end{equation}

\noindent where $T^{N,W}_{mb}$ denotes the number of DRs using MB strategy and $T^{N,W}_{\text{TNS}}$ denotes the number of DRs using basic TNS. Moreover, using smaller bank sizes usually improve the operating frequency and achieve faster sorting speed. MB strategy solves the scalability problem for sorting practical large-scale dataset partitioned based on numbers and stored in different memory banks. \figurename~\ref{fig:ca_strategies}a-b and \figurename~\ref{fig:ca_strategies}f-g show the operation flow of MB strategy and performance comparisons. MB strategy outperforms ASIC merge sorter by nearly $3\times$ in speed across different benchmarks and the speedup goes up with more banks before reaching an optimal point. Detailed analysis are provided in Supplementary Section~\ref{MB_Strategy} and \ref{CA_TNS_Hardware}.

\subsubsection{Bit-Slice Strategy}\label{subsubsec2}

Data numbers are usually represented by certain quantization formats using $W$ bits, such as unsigned fixed-point numbers or floating-point numbers. Some classical choices of $W$ can be 8 bits or 32 bits depending on the accuracy requirements of computing tasks. Although TNS with MB strategy can improve sorting speed by partitioning the dataset based on numbers, there is still a possibility to introduce higher level of parallelism for further speedup. Here, we introduce the bit-slice (BS) strategy, which partitions the dataset into several parts based on digits and stores them into $n$ different memristor arrays as shown in \figurename~\ref{fig:ca_strategies}c. Similar to MB strategy, each memristor bank can reuse the periphery and behave as an independent sub-sorter for $N$ numbers with data precision $W_{bsi}$ bits, where $W_{bsi}$ follows \eqref{bs_eq} and $n$ sub-sorters run successively from upper digits to lower digits:

\begin{equation}
\label{bs_eq}
\sum_{i = 1}^{n}W_{bsi} = W  
\end{equation}

Upper digits are generally more significant than lower digits; therefore, if we locate min/max values at sub-sorter storing upper digits, we do not need to continue the min/max search at other sub-sorters storing lower digits. Otherwise, if we can not locate min/max values at sub-sorter storing upper digits, i.e., there are multiple numbers with identical upper digits, the status of these survival numbers needs to be sent to sub-sorters storing lower bits to continue the min/max search. Meanwhile, the sub-sorter storing upper digits can start next min/max searcg iteration, creating a pipelined scheduling between successive sub-sorters for different min/max search iterations and accelerating the sorting process. $n-1$ cross-array processors configured as bit-slice (BS) mode are instantiated to transfer survival number information from upper digits sub-sorters to lower digits sub-sorters successively. The number of DRs using BS strategy can be reduced to approximately the maximum number of DRs on sub-sorters as shown in \eqref{bs_sub_latency}:

\begin{equation}
\label{bs_sub_latency}
T^{N,W}_{bs} \approx \text{max}_{i}(T^{N,W_{bsi}}_{\text{TNS}}), \text{ } i = 1\rightarrow n
\end{equation}

\noindent where $T^{N,W}_{bs}$ denotes the number of DRs using BS strategy and $T^{N,W_{bsi}}_{\text{TNS}}$ denotes the number of DRs on the $i$-th sub-sorter. \figurename~\ref{fig:ca_strategies}c-d and \figurename~\ref{fig:ca_strategies}f-g show the operation flow and performance of BS strategy. BS strategy outperforms ASIC merge sorter by up to $7\times$ in speed across different benchmarks. Detailed implementations and performance analysis are provided in Supplementary Section~\ref{BS_Strategy} and \ref{CA_TNS_Hardware}.

\subsubsection{Multi-Level Strategy}\label{subsubsec3}

Multi-level capability is one of the key advantages of memristors, where devices can be programmed to many different conductance states by applying different designed pulses \cite{yao2020fully,lastras2021ratio}. For sorting purpose, multi-level devices enable in-device parallelism that can retrieve more information from each DR. To support multi-level (ML) strategy, we extend state controller and digit processor to support $n$-bit multi-level (ML-$n$-bit) devices. For example, the DR results using ML-$2$-bit devices may include different combinations of 11's, 10's, 01's and 00's, among which only the smallest ones need to be kept for number exclusions. On the other hand, only $\lceil N \times W/n \rceil$ devices are needed to store all the data and the periphery only needs length-$\lceil W/n \rceil$ digit registers and LIFOs; therefore, the number of DRs can be reduced to \eqref{ml_sub_latency}:

\begin{equation}
\label{ml_sub_latency}
T^{N,W}_{ml} \approx T^{N, \lceil W/n \rceil}_{\text{TNS}}
\end{equation}

\noindent where $T^{N,W}_{ml}$ denotes the number of DRs using ML strategy and $T^{N, \lceil W/n \rceil}_{\text{TNS}}$ denotes the number of DRs when sorting data with precision $\lceil W/n \rceil$ using basic TNS. 

Here we demonstrate our SIM system supporting up to ML-$3$-bit with help of write-verify rules as shown in \figurename~\ref{fig:write_verify}. The write-verify scheme is carried out iteratively using designed pulses based on target conductance $G^{\text{target}}_{i}$ and conductance error tolerances $\Delta G_i$, where $i = 1\rightarrow 2^n$. Using the same $\Delta G_i$ for $i = 1 \rightarrow 2^n$, we observe that the programming effort (measured by the number of pulses needed to reach $G^{\text{target}}_{i}\pm \Delta G_i$) grows and gets saturated with $G^{\text{target}}_{i}$ increases as shown in Supplementary \figurename~\ref{fig:fixerror}. Hence we use larger $\Delta G_i$ for larger target $G^{\text{target}}_i$ aiming to lower the program failure rate (PFR) and reduce the possible conductance overlaps between adjacent conductance levels introduced by ML strategy. Detailed multi-level device measurements and performance evaluations are given in Supplementary Section~\ref{MLC} and \ref{ML_Strategy}. On the other hand, even with write-verify methodology mentioned above, memristor devices may still induce bit errors due to overlapped conductance states. Therefore, we propose another approach for pseudo multi-level strategy, which uses multiple binary devices to simulate a multi-level device. For example, to simulate ML-$2$-bit cells, we can store odd and even bits of data in separate memristor arrays. The DRs of odd sub-sorter and even sub-sorter can be processed together, reusing the same periphery that process ML-$2$-bit DRs. In this way, the number of devices is kept unchanged as basic TNS, but complex ADCs can be replaced by SAs and sorting speed can be improved while maintaining accuracy of non-ML devices. \figurename~\ref{fig:ca_strategies}e demonstrates the two approaches in realizing ML strategy and \figurename~\ref{fig:ca_strategies}f-g present the speedup and energy efficiency enhancements when using ML strategies.

\subsection{Memristor-based SIM System Design}\label{subsec3}

We design an end-to-end fast and reconfigurable hardware and software co-designed sorting system to carry out experiments based on memristor arrays. Our system consists of four modules: 1T1R memristor array chip, PCB board with periphery, FPGAs and control PC (\figurename~\ref{fig:introduction}c and \figurename~\ref{fig:system}). The memristor array chip has 32$\times$32 1T1R crossbars of TiN/TaOx/HfO2/TiN memristors using 180nm technology. We test a total of 800 cycles on 100 different devices and the experimental results are shown in \figurename~\ref{fig:architecture}a-g. Without adopting write-verify methodology, the memristors have a very high on-off ratio approximately 16.14$\times$ (\figurename~\ref{fig:architecture}c).  We design the write/read circuitry on the PCB for data writing and DR operation in TNS/CA-TNS. The FPGA is used to control and receive DR data from the chip to implement the logic functions of TNS/CA-TNS. The end-to-end system employ a control PC that communicates with FPGAs to configure dataset elements, data precision, number representations and TNS/CA-TNS parameters while FPGA sends the sorting result to PC for further computing, as shown in \figurename~\ref{fig:system}. Detailed implementations including periphery, PCB and FPGAs design are provided in Supplementary Section~\ref{PCB}.

\section{Experiments and Results}\label{sec3}

We first adopt our SIM system for a representative real-world sorting problem, i.e., shortest path search using the Dijkstra's algorithm\cite{dijkstra2022note}, where 16-bit floating-point numbers are used. Furthermore, we apply our SIM techniques to neural network (PointNet++ \cite{qi2017pointnet++}) inference with run-time tunable sparsity using 8-bit fixed-point numbers. Memristor-based SIM techniques present high compatibility of integrating with other CIM techniques such as matrix-vector multiplications, enabling in-situ pruning for improved system performance.

\subsection{Shortest Path Search with Dijkstra's Algorithm}\label{subsec3.1}

\begin{figure}[hbt!]
    \centering
    \includegraphics[width = 1\linewidth]{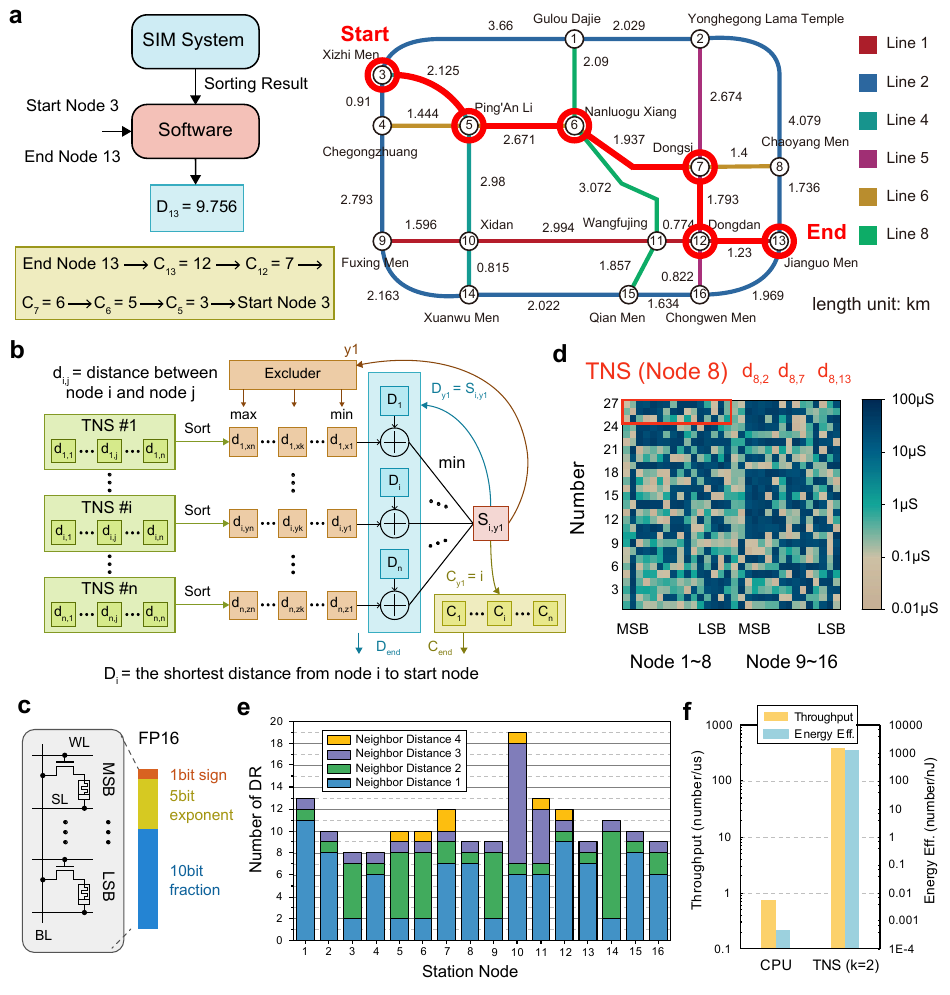}
    \caption{\textbf{TNS experiment for shortest path search.} \textbf{a,} Details of 16 Beijing subway stations for testing the memristor-based Dijstra's algorithm using TNS. \textbf{b,} Diagram of Dijstra's algorithm for shortest path search. \textbf{c,} Mapping methods for half-precision floating point numbers. \textbf{d,} Conductivity distribution map that stores all the neighboring distances for 16 stations. \textbf{e,} Numbers of DR for sorting the neighbor distances of 16 station nodes with record LIFO size 2 ($k=2$) in basic TNS. \textbf{f,} Average throughput and energy efficiency of sorting 16 station nodes in Dijkstra's algorithm using CPU and proposed SIM with TNS ($k=2$).}
    \label{fig:dijstra}
\end{figure}

Dijkstra's algorithm iteratively selects the nearest neighbor at a time and adds it to previous shortest path. It has been widely used in scenarios such as multi-source shortest path problem \cite{dijkstra2022note}, minimum spanning tree problem \cite{mst1976} and a variety of problems in image processing \cite{lin2019dijkstra,wang2017greedy,vicente2008graph,lempitsky2009image,sinop2007seeded}, robot navigation \cite{ab2020comparative,li2021openstreetmap,zhang2015localization,vasquez2014inverse} and traffic analysis \cite{zheng2020determinants}, etc. Suppose we have a graph consisting of multiple nodes and the edges between nodes have associated cost functions. The algorithm works by maintaining a set of nodes for which the shortest path from the source node is known. Initially, this set contains only the source node. By iterating through the nodes in the graph, the node closest to the source that has not yet been processed is selected, and its distance from the source is determined. This distance is then used to update the distance of each of its unvisited neighbors. The process continues until all nodes have been processed.
We conduct an experimental demonstration on the shortest path search problem between subway stations in Beijing. The distances between each subway station node and all its neighboring nodes are represented by 16-bit floating-point numbers. A 16-bit floating-point number can be stored in 16 1T1R cells, 1 for sign bit, 5 for exponent bits and 10 for fraction bits (\figurename~\ref{fig:dijstra}c). The high and low resistance states of each memristor device represent 0's and 1's, respectively. Following the TNS for floating-point numbers, for each node in the graph, we utilize memristor-based SIM system to sort all distances from its neighboring nodes. The min value of each TNS is added to $D_{i}$ (the shortest distance from node $i$ to input node) to calculate $S_{i,y1}$, which is then used to update $D_{i}$ and $C_{i}$ (the predecessor of node $i$). The process repeatedly excludes node $y1$ corresponding to $S_{i,y1}$ from previous sorting results until reaching the output node (\figurename~\ref{fig:dijstra}b, \figurename~\ref{fig:dijstra_flow}). We pick 16 subway stations from 6 subway lines in Beijing for experimental purpose, where each node has 3 or 4 neighboring nodes as shown in \figurename~\ref{fig:dijstra}a. We divide our 1T1R array into two parts, one for station nodes 1 to 8 and the other for 9 to 16, storing a total of 54 distances in 16-bit floating-point numbers. The conductance distribution diagram of the memristor devices in the 1T1R array is shown in \figurename~\ref{fig:dijstra}d. Since the dataset is small, there is no need to adopt cross-array strategies mentioned in previous section. We use basic TNS and measure the number of cycles required for sorting different station nodes with TNS parameter $k=2$. The results are shown in \figurename~\ref{fig:dijstra}e and it takes approximately 3 DRs to sort a number on average. \figurename~\ref{fig:dijstra}f demonstrates that our TNS achieves an average sorting throughput and energy efficiency of nearly 400 numbers/$\mu s$ and 1300 numbers/nJ, respectively, outperforming CPU running the same task by over three orders or magnitudes.  

\subsection{Neural Network with Run-time Tunable Sparsity}\label{subsec3.2}

\begin{figure}[hbt!]
    \centering
    \includegraphics[width = 1\linewidth]{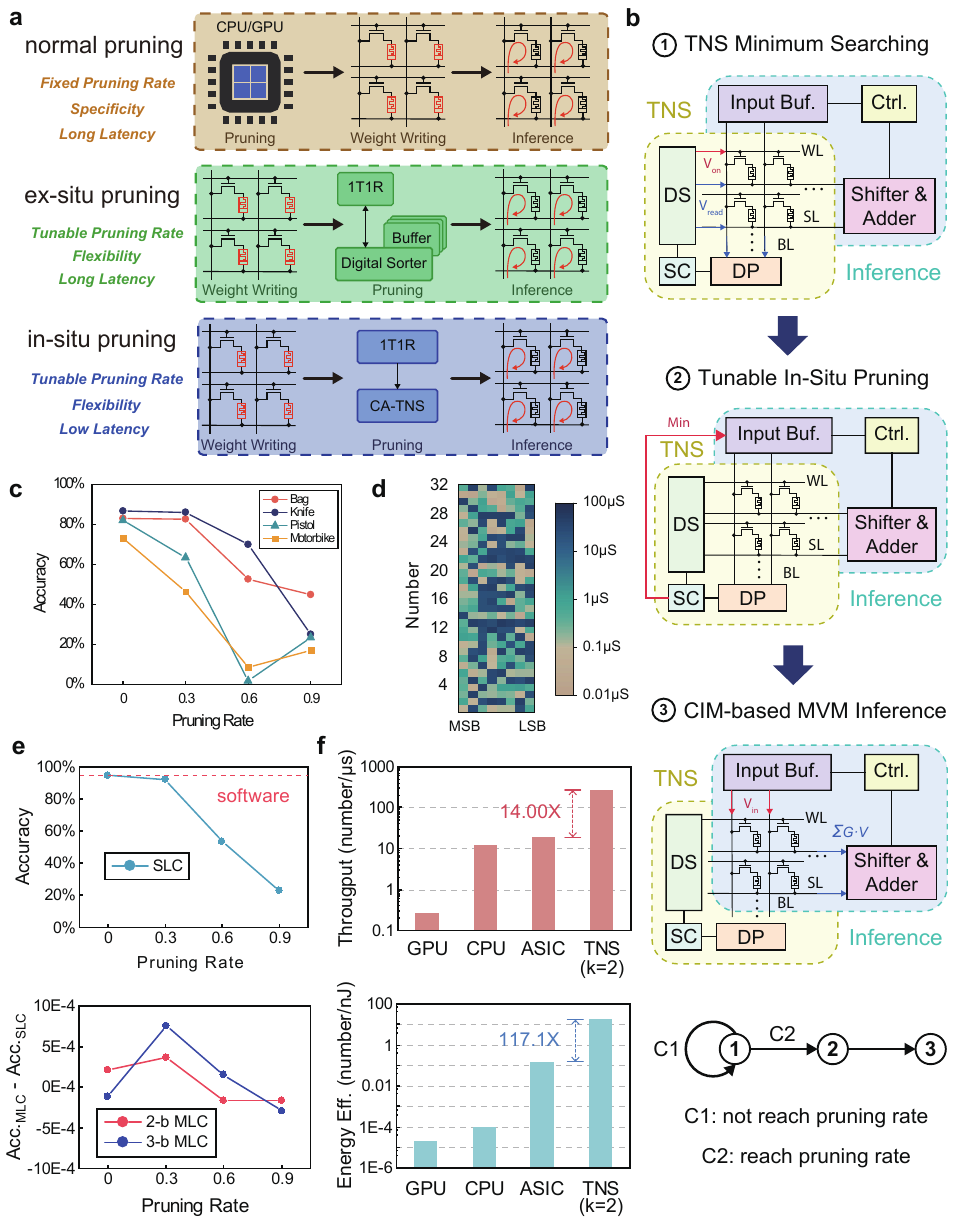}
    \caption{\textbf{In-situ pruning for PointNet++.} \textbf{a,} Normal, ex-situ and in-situ pruning. \textbf{b,} Operation flow diagram. \textbf{c,} Accuracy for four classical objects in PointNet++. \textbf{d,} Example conductance distribution map for a 8-bit fixed-point batch normalization (BN) layer. \textbf{e,} Accuracy with multi-level devices. \textbf{f,} Throughput and energy efficiency using CPU/GPU, ASIC and TNS ($k=2$) with 30\% pruning rate.}
    \label{fig:pointnet}
\end{figure}

State-of-the-art neural networks have massive weight parameters but their significance are different; hence, sparsity is exploited to eliminate unnecessary computing for targeted tasks \cite{mahmoud2020tensordash, raihan2020sparse, yousefzadeh2021training} without affecting model performance. Run-time tunable sparsity \cite{raihan2020sparse} is adopted to support diversified use cases using one trained model. \figurename~\ref{fig:pointnet}a demonstrates various CIM approaches to realize pruning. Normal pruning methods pre-prune the weights by CPUs/GPUs before writing them to CIM arrays and can only support a fixed pruning rate. Ex-situ pruning extends normal pruning to support tunable pruning rate, but it requires additional digital sorters and relevant buffers to update the corresponding inputs to CIM arrays, incurring extra latency and area/energy cost (\figurename~\ref{fig:pointnet}a). \figurename~\ref{fig:pointnet}c shows the motivations behind run-time tunable sparsity, where different pruning percentages are required to achieve desired inference accuracy for different types of classification tasks. Meanwhile, state-of-the-art CIM techniques are mainly designed for matrix-vector multiplications (MVM) in neural network inference, lacking supports for run-time tunable sparsity.

Here we integrate our TNS-based SIM techniques with MVM CIM for in-situ pruning and inference as shown in \figurename~\ref{fig:pointnet}b, demonstrating their performance enhancing capabilities on representative neural network naming PointNet++ \cite{qi2017pointnet++}. The dominant computations in PointNet++ are convolution and batch normalization layers with trainable weights that can be stored in our 1T1R memristor arrays. We use TNS to sort the weights of each layer and identify the $p\%$ weights with the smallest absolute values. Each time we locate a min weight, we store its address and mask the corresponding input to 0 to discard it in subsequent MVM inference (Supplementary Section~\ref{Details_PointNet}). Here we first apply read voltages on select lines (SL) and read the bit lines (BL) to complete the DR operations in TNS. In later inference, we apply input voltages on BL and sense the MVM results on SL (\figurename~\ref{fig:pointnet}b). For experimental demonstration, we select a batch normalization layer with a total of 32 weights and map them to 8-bit fixed-point numbers on memristor array. The conductance distribution diagram of the devices in the array is shown in \figurename~\ref{fig:pointnet}d. We also record a demonstration video to illustrate the sorting process that locates $30\%$ of the weights in that layer with smallest absolute values (Supplementary Video 2).

In order to further improve storage density and sorting speed, we also utilize multi-level devices and experimentally test inference performance with different multi-level choices. Multi-level cells may introduce higher error rates in computing; however, neural network inference are usually nonsensitive to small noise in their model weights \cite{huang2023nonvolatile}. By selecting appropriate target conductance with proposed write-verify scheme, our memristors can be written to eight conductance states. The cumulative distribution function (CDF) of the eight conductance states is shown in \figurename~\ref{fig:architecture}e. By using single-level (SL) cells, 2-bit multi-level (ML-$2$-bit) cells and 3-bit multi-level (ML-$3$-bit) cells to store the 8-bit weights, a design trade off exists between inference accuracy and storage density. We investigate three extreme cases of hybrid precision mapping (8 SL cells, 4 ML-$2$-bit, and 1 ML-$2$-bit plus 2 ML-$3$-bit to present a 8-bit weight). Based on experimental testing results, the average programming failure rate (PFR) of our multi-level programming using proposed write-verify scheme is 1.224\% across the 8 conductance states (the PFR is measured as the probability of a device failed in converging at target conductance plus/minus associated conductance error tolerance). Detailed PFRs for different conductance levels are provided in \ref{MLC}). We further investigate the recognition accuracy using our memristor-based SIM system at different pruning percentages (\figurename~\ref{fig:pointnet}e). The accuracy impacts of multi-level cells are almost negligible compared with single-level cells. Since PFR often leads to bit error, we study the effects of bit error rate (BER) on the recognition accuracy and show that PointNet++ achieves high recognition accuracy even at nearly 20\% of BER (\figurename~\ref{fig:ber}). Compared to GPU-, CPU- and ASIC-based systems running the same inference with tunable sparsity, our TNS-based SIM system outperforms by more than 14.00$\times$ and 117.1$\times$ in throughput (number/$\mu s$) and energy efficiency (number/nJ), respectively, as shown in \figurename~\ref{fig:pointnet}f.

\section{Conclusions}\label{sec4}

In this article, we report an end-to-end memristor-based hardware and software co-designed SIM system using TaOx and HfO2 based memristor array chips, periphery circuits and FPGAs, aiming to completely eliminate comparison units and their relevant data transfers using iterative min/max search based sorting. We develop tree node skipping (TNS) methods to optimize the comparison-free sorting and also consider special cases such as repeating numbers or last numbers in the dataset. Furthermore, we extend basic TNS to cross-array TNS (CA-TNS) strategies: multi-bank for improving number-based scalability, bit-slice for improving digit-level parallelism, and multi-level for improving storage density as well as sorting speed. Compared with CPU/GPU-based or ASIC-based sorting systems, experimental results show that our memristor-based SIM system greatly improves sorting speed, energy efficiency and area cost across five classical sorting datasets by $3.32\times \sim 7.70 \times$, $6.23 \times \sim 183.5 \times$ and $2.23 \times \sim 7.43 \times$, respectively. The proposed SIM techniques also support variable data quantity, data types and data precision. Applying such SIM system to two representative real-world applications, shortest path search using Dijkstra's algorithm and neural network (PointNet++) inference with run-time tunable sparsity, the experimental results demonstrate strong capability in solving practical sorting problem and excellent compatibility in integrating with conventional in-memory MVMs. Memristor-based SIM system using TNS/CA-TNS offers high speed, high energy efficiency and high scalability with low area cost, supporting variable data types and compatibility with other CIM techniques. The comparison-free concept and the TNS/CA-TNS strategies demonstrate high potential in future sorting system design based on memristor devices.

\section{Methods}\label{sec11}

\subsection{1T1R Array Chip Fabrication}\label{fabrication}

The 1T1R (One Transistor One RRAM) crossbar array was taped out based on standard 180 nm CMOS technology. The transistors fabricated in the FEOL (Front End of Line) serve as the select units of the 1T1R cells. Five layers of metal were then used for interconnection purpose. After the last W via formation followed by CMP (Chemical Mechanical Polishing) process, the wafers were transferred to an RRAM production line for subsequent RRAM fabrication processes. The exposed W vias on the CMOS substrates were first cleaned with argon plasma to remove the native oxide, after which the RRAM cells with HfO2 as switch layer were deposited on top of the W vias. The top and bottom electrode metal TiN were deposited by sputtering and the dielectric layer of HfO2 was deposited by ALD (Atomic Layer Deposition). Another via was formed and the metal layer was grown to finalize the entire process.

\subsection{Conductance Programming with Write-Verify}

We experimentally demonstrate programming with the write-verify scheme supporting up to 8 conductance states in our fabricated 1T1R array chip. The memristor array chip combined with PCB and FPGAs are connected together for demonstration of the TNS-based SIM system. For basic TNS, multi-bank and bit-slice strategies, we use direct-current (DC) module of Agilent B1500A to program the memristor devices to LRS or HRS state. Under DC writing conditions, all of the memristor devices can reach the desired binary conductance states. For multi-level strategy, to reliably differentiate 8 conductance states, we gradually program the conductance values by applying designed voltage pulses through Agilent B1530 pulse generator. Accordingly, we program the memristor conductance into a defined range from $G^{\text{target}}_i \pm \Delta G_i$ ($i = 1\rightarrow 8$) using a closed-loop writing method implemented using a script in LabVIEW 2010: 

\renewcommand\labelenumi{(\theenumi)}
\begin{enumerate}[\indent(1)]
    \item If $G^{read}_i < (G^{\text{target}}_i - \Delta G_i)$, a SET pulse would be applied on the TE of memristor device;
    \item Otherwise, if $G^{read}_i > (G^{\text{target}}_i + \Delta G_i)$, a RESET pulse would be applied on the BE of memristor device (\figurename~\ref{fig:write_verify}a).
\end{enumerate}

The experimental results demonstrate that most of the programmed memristive devices are located within the defined conductance range (\figurename~\ref{fig:8level}). To allow the device to reach the target conductance faster while maintaining a tolerable programming failure rate (PFR), we choose $\Delta G_i$ proportional to $G^{\text{target}}_i$ (\figurename~\ref{fig:write_verify}b). With such conductance selections and write-verify rules, we need an average of 13.95 pulses to program the device to the target conductance with an average PFR = 1.224\% across 8 different conductance states (\figurename~\ref{fig:ini_count}b).

\subsection{System Integration for End-to-End Demonstration}

The memristor-based SIM system is based on software and harware co-design using 32×32 1T1R array chip, a custom-designed PCB, Altera DE2-115 FPGA and PC. The custom PCB includes decoders and analog switches to select devices for writing and sense amplifiers as well as analog-to-digital converters for TNS digit read (DR) programmed as single-level or multi-level devices. Other logic for TNS, such as state registers and LIFOs, are deployed on FPGAs. We use Agilent B1500A and FPGAs to program our fabricated memristor devices. The sorting results are output by the logic on FPGAs and sent to software for subsequent processing. Readers can refer to \ref{PCB} for detailed information on the integration of the system.

\subsection{Sorting Datasets and Performance Evaluations}

We evaluate proposed SIM performance using five widely-used sorting benchmark (random, normal, clustered, Kruskal's and MapReduce) with low (8-bit) and high (32-bit) precision unsigned fixed-point numbers, respectively. For low-precision 8-bit datasets, the randomly-distributed dataset ranges from 0 to $2^{8}-1$, the normal-distributed dataset has a mean of $2^{7}$ and a standard deviation of $2^{7}/3$, and the clustered-distributed dataset has 2 clusters centered at $100$ and $200$ with identical standard deviation of $10$. For high-precision 32-bit datasets, the parameters of the three statistically-distributed datasets are modified accordingly, where the randomly-distributed dataset ranges from 0 to $2^{32}-1$, the mean and standard deviation of the normal-distributed dataset change to $2^{31}$ and $2^{31}/3$, respectively, and the 2 clusters in the clustered-distributed dataset center at $2^{15}$ and $2^{25}$ with identical standard deviation of $2^{13}$. In addition to the three statistically-distributed datasets mentioned above, we also evaluate our SIM performance on two sorting benchmark datasets (from Kruskal's and MapReduce) quantized in 32-bit unsigned fix-point numbers. The speed, energy efficiency and area cost evaluations of our memristor-based SIM system are provided in \ref{Performance_Comp}.

\section{Data availability}

The experimental data associated with this article are available upon requests to the corresponding authors.

\backmatter

\newpage 
\noindent\textbf{Acknowledgements}

\noindent\textbf{Funding:}

This work was supported by the National Natural Science Foundation of China (61925401, 92064004, 61927901, 8206100486, 92164302) and the 111 Project (B18001). Y.Y. acknowledges support from the Fok Ying-Tong Education Foundation and the Tencent Foundation through the XPLORER PRIZE.

\noindent\textbf{Conflict of interest/Competing interests:} 

The authors declare no conflict of interest/competing interests.

\noindent\textbf{Availability of data and materials:}

Data and materials are available upon requests.

\noindent\textbf{Code availability:} 

Codes are available upon requests.

\noindent\textbf{Authors' contributions:}

Conceptualization: Lianfeng Yu, Yaoyu Tao

Methodology: Lianfeng Yu, Yaoyu Tao, Teng Zhang

Investigation: Lianfeng Yu, Yaoyu Tao, Zeyu Wang, Xile Wang, Zelun Pan, Bowen Wang, Teng Zhang, Yihang Zhu, Jiaxin Liu, Yuqi Li

Visualization: Lianfeng Yu, Yaoyu Tao, Zeyu Wang, Xile Wang

Supervision: Yaoyu Tao, Bonan Yan, Yuchao Yang

Writing—original draft: Lianfeng Yu, Yaoyu Tao

Writing—review \& editing: Lianfeng Yu, Yaoyu Tao, Yuchao Yang




\newpage 
\bibliography{sn-bibliography}


\newpage
\bmhead{Supplementary information}

\renewcommand{\thesection}{S\arabic{section}}
\renewcommand{\thefigure}{S\arabic{figure}}
\renewcommand{\thetable}{S\arabic{table}}
\renewcommand{\thealgorithm}{S\arabic{algorithm}}
\setcounter{section}{0}
\setcounter{figure}{0}
\setcounter{table}{0}
\setcounter{algorithm}{0}

\section{Binary Memristors Programming}\label{SLC}

\begin{figure}[hbt!]
    \centering
    \includegraphics[width = 1\linewidth]{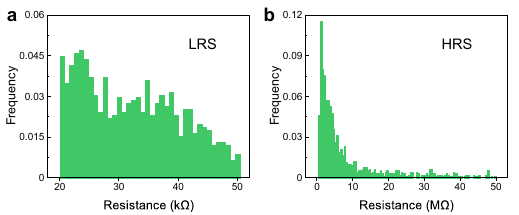}
    \caption{The conductance distribution of our memristors under DC scan for \textbf{a,} LRS and \textbf{b,} HRS.}
    \label{fig:2level}
\end{figure}

We program the memristors to binary LRS and HRS using DC scanning. With a set voltage of 2V and a reset voltage of 2.4V, our devices show good switching characteristics. There is no programming error and the lowest ON/OFF switching ratio is 16.14$\times$. The programmed resistance distribution is shown in \figurename~\ref{fig:2level}.

\newpage
\section{Multi-level Memristors Programming}\label{MLC}

\begin{figure}[hbt!]
    \centering
    \includegraphics[width = 1\linewidth]{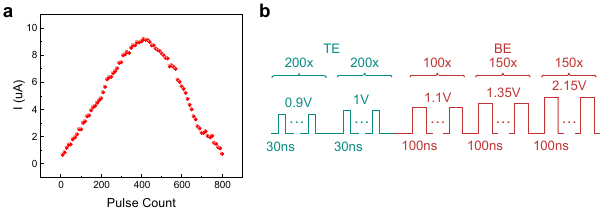}
    \caption{\textbf{a,} The conductance states of memristors with number of designed pulses. \textbf{b,} Design for a total of 800 pulses to increase and decrease the conductance of our memristors.}
    \label{fig:ltpd}
\end{figure}

The first 400 pulses are applied to the top electrode (TE) of the devices and the last 400 pulses are applied to the bottom electrode (BE). The average results of applying stimuli to 100 memristor devices are shown in \figurename~\ref{fig:ltpd}a. \figurename~\ref{fig:ltpd}b shows detailed information about the 800 designed pulses. Our devices exhibit good linearity and potential for multi-level programming.

\newpage
\begin{figure}[hbt!]
    \centering
    \includegraphics[width = 0.8\linewidth]{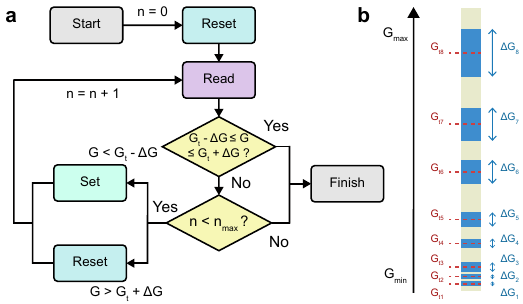}
    \caption{\textbf{a,} Flowchart of our write-verify rules. \textbf{b,} Target conductance states $G_{ti}$ for 8-level devices with error tolerances $\Delta G_i$, where $i = 1\rightarrow n$.}
    \label{fig:write_verify}
\end{figure}

\begin{figure}[hbt!]
    \centering
    \includegraphics[width = 0.5\linewidth]{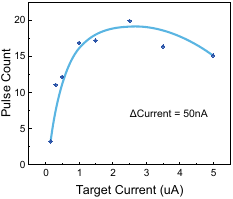}
    \caption{\textbf{Number of pulses required to program a typical device to 8 conductance states under a fixed current offset $\Delta I = 50nA$}}
    \label{fig:fixerror}
\end{figure}

We design a write-verify scheme to enable fast and accurate programming to the target conductance states (\figurename~\ref{fig:write_verify}a). Firstly, we reset the memristor device to the high resistance state. Secondly, an iterative programming procedure starts, where each time we read the device conductance to determine whether the programmed conductance is close enough to the target conductance ($G_{ti}-\Delta G_i \le G_{read} \le G_{ti}+\Delta G_i$). If so, we finish the writing process; otherwise, if the programmed conductance is out of target range and the maximum number of programming cycles we set ($n_{max}$) is not reached, we apply a SET pulse or a RESET pulse depending on the readout conductance. Finally, the device conductance is read again for the next programming iteration. Considering conductance overlap and programming efforts, we finally choose 8 resistance states (\figurename~\ref{fig:write_verify}b) based on non-linear target conductance. Through experiments, we also find that when using the same conductance error tolerances ($\Delta G_i$) for different conductance states, the programming efforts measured by number of pulses increase and then get saturated with target $G_{ti}$. \figurename~\ref{fig:fixerror} shows the average pulse counts required to write to different conductance states for fixed $\Delta G_i$. Near the LRS regime, the number of pulses decreases sharply because the conducting filaments are more stable. In summary, we choose $G_{ti}$ based on non-linear target conductance and $\Delta G_i$ proportional to $G_{ti}$ to maintain acceptable convergence time reaching target conductance while still keeping conductance overlap small.

\newpage

\begin{figure}[hbt!]
    \centering
    \includegraphics[width = 1\linewidth]{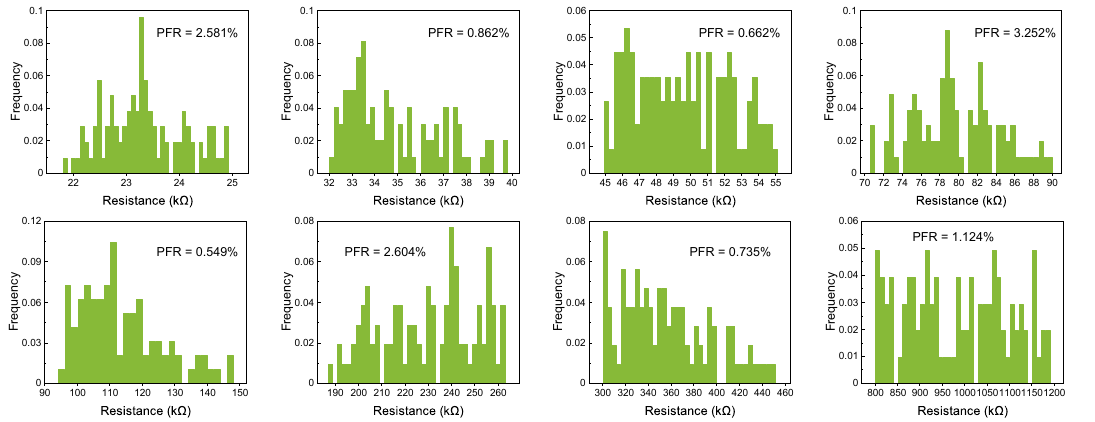}
    \caption{The 8-level resistance distribution of our memristors with write-verify rules.}
    \label{fig:8level}
\end{figure}

The resistance distribution of the devices in the 8 conductance states is shown in \figurename~\ref{fig:8level}. We also label the average PFR when writing the devices to each conductance state.

\newpage
\section{Detailed Operations of BTS}\label{Details_BTS}

\begin{figure}[hbt!]
    \centering
    \includegraphics[width = 1\linewidth]{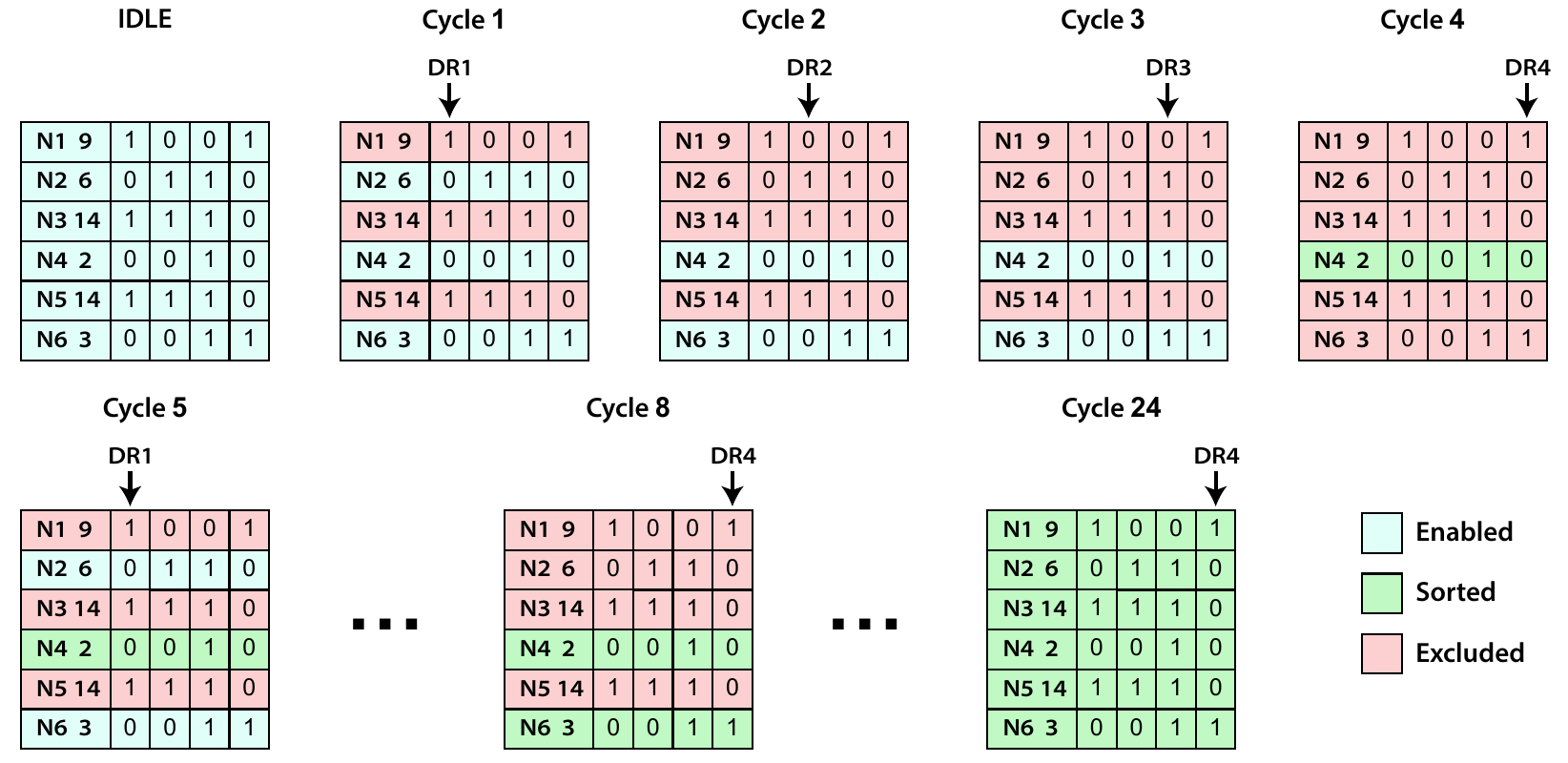}
    \caption{\textbf{An example of sorting six unsigned 4-bit fixed-point numbers with BTS.}}
    \label{fig:exampleBTS}
\end{figure}

We use a more detailed example of sorting six 4-bit unsigned fixed-point numbers \{2, 3, 9, 6, 14, 14\} to illustrate BTS process (\figurename~\ref{fig:exampleBTS}). We start DR1 from MSB, and in cycle 1, we find that 0's and 1's exist simultaneously in DR results. Here DR$i$ denotes digit read at $i$-th column. Therefore, we exclude the three numbers 9, 14, and 14 with DR results 1's. Using the same method, we go to the next bit and exclude 6 in cycle 2. In cycle 3, because the DR results of remaining enabled numbers are all 1's, we do not need to perform number exclusions (NE). In cycle 4, since LSB has already been reached at this point, we exclude 3 and find the minimum value of 2. After the first iteration, we return to MSB and repeat the process in four cycles. Until cycle 24, we finally complete the sorting of a total of six numbers\cite{yu2022fast}. 

\newpage
\section{Detailed Operations of TNS}\label{Details_TNS}

\begin{figure}[hbt!]
    \centering
    \includegraphics[width = 0.8\linewidth]{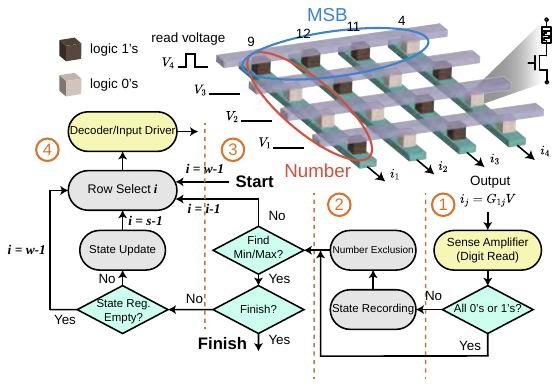}
    \caption{\textbf{TNS flowchart divided into four parts.}}
    \label{fig:TNSflow}
\end{figure}

The detailed workflow of TNS is shown in \figurename~\ref{fig:TNSflow}, which can be described as follows:  

\renewcommand\labelenumi{(\theenumi)}
\begin{enumerate}[\indent(1)]
    \item For $i$-th digit, read voltages are applied on memristor array and DR results are sensed. If DR results are all 0's or all 1's, the process jumps to min/max check; otherwise, the process continues for state recording (SR) and number exclusion (NE);
    \item SR records the tree node information with digit index and number status. Each number in the dataset holds a status of valid (available for min/max search), excluded (discarded for min/max search) or sorted. Then, NE excludes enabled numbers with DR results 1's for min search or DR results 0's for max search, respectively;
    \item Upon completion of number exclusions, we examine whether min/max value of current search iteration is determined, which can happen in two scenarios: TNS reaches the LSB, or there is only one enabled number left before TNS reaches LSB. If neither scenarios occur, the process jumps to DR of the $(i-1)$-th digit;
    \item If min/max value is found but dataset sorting is not completed, we reload the tree node stored in the state registers if they are nonempty (Supplementary Section~\ref{Details_TNS}). State update is carried out according to the reloaded tree node, based on which we can start DR from an intermediate digit $s$-1 instead of starting from MSB.
\end{enumerate}

\noindent TNS enables DR tree traversal starting from an intermediate tree node (other than MSB) and ending at another intermediate tree node (before reaching LSB), reducing the SIM latency from $O(N\times W)$ to $O(N+W)$ in the best case.

\begin{figure}[hbt!]
    \centering
    \includegraphics[width = 1\linewidth]{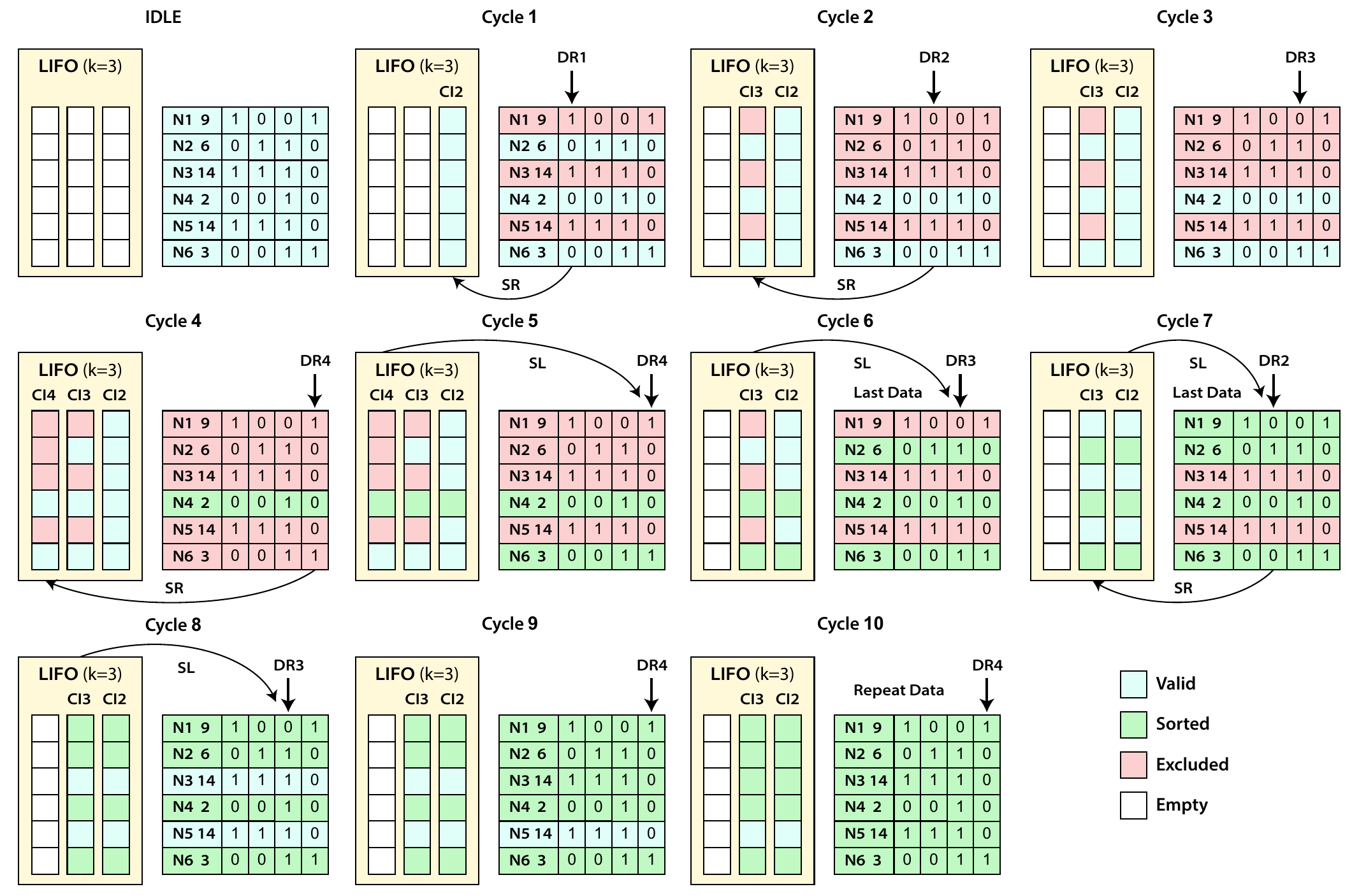}
    \caption{\textbf{The flow chart when sorting the same six unsigned 4-bit fixed-point numbers as Supplementary Section~\ref{Details_BTS} with TNS.} }
    \label{fig:exampleTNS}
\end{figure}

\begin{figure}[hbt!]
    \centering
    \includegraphics[width = 1\linewidth]{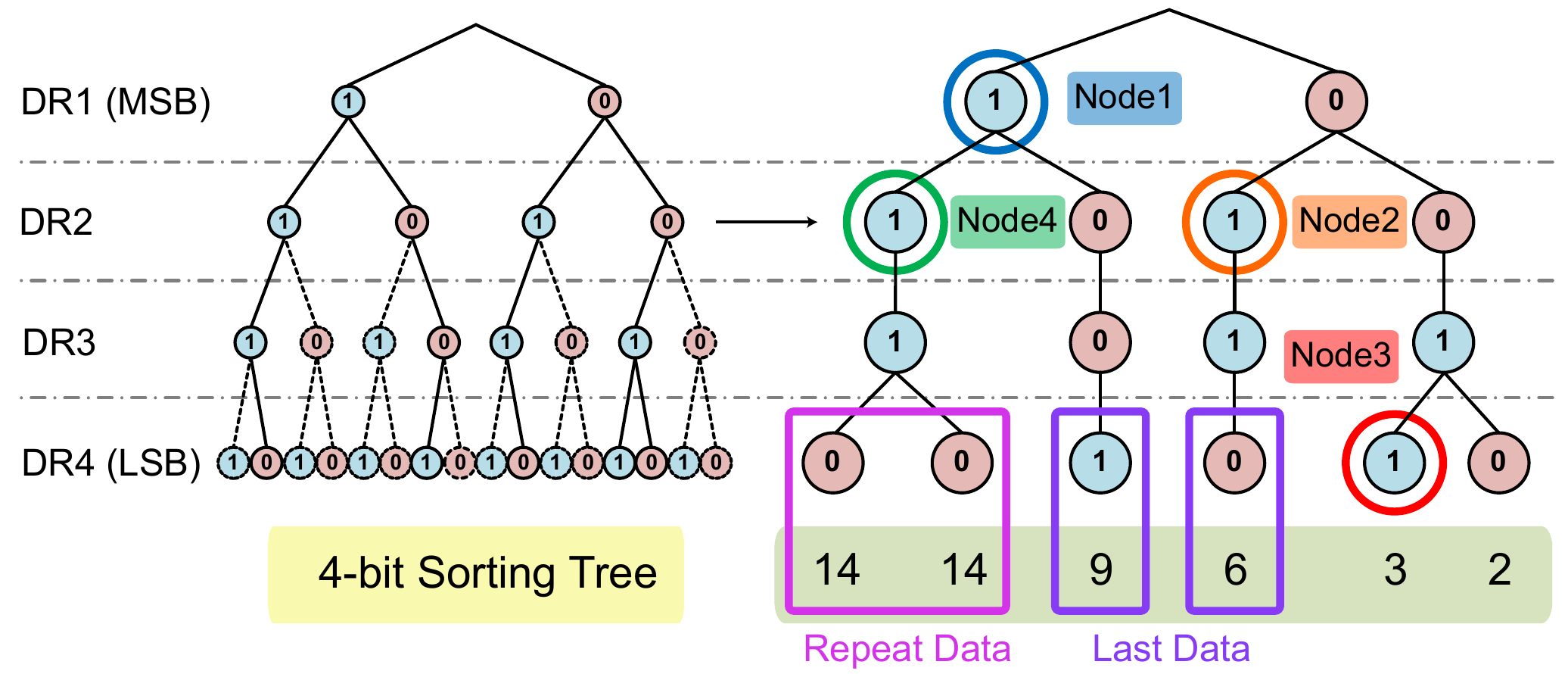}
    \caption{\textbf{The 4-bit DR tree and recorded nodes corresponding to the example in \figurename~\ref{fig:exampleTNS}.}}
    \label{fig:TNStree}
\end{figure}

Suppose LIFO size is 3 ($k=3$) and we use TNS to sort the six unsigned 4-bit fixed-point numbers as Supplementary Section~\ref{Details_BTS} (\figurename~\ref{fig:exampleTNS}). The initial DR1 starts from MSB at 1st column. Here DR$i$ denotes digit read at $i$-th column. In cycle 1, DR results include both 0's and 1's. We record the information of the tree nodes in LIFO with state recording (SR), including the DR results and the number status. Specifically, we record number status at column index 2 (CI2) in LIFO (indicating the next column to read) and the number status where all numbers that are valid when executing DR1 (labeled as all blue in \figurename~\ref{fig:exampleTNS}). This is because each tree node has only two sub-trees; after we complete one sub-tree, we can go directly to the other sub-tree. When we reload this node in future cycle (cycle 7), all the numbers corresponding to its sub-tree nodes with DR results 0's have been sorted, so we can directly go to DR of the next digit. After SR, we carry out number exclusion (NE) where numbers with DR results 1's are excluded and labeled as red. Similarly in cycle 2 and cycle 3, we record the tree node states (CI3 and CI4 and their corresponding number status) in LIFOs and then exclude 6 and 3 successively. We locate the min value 2 at cycle 4 when there's only one valid number left in the array and label it as sorted. Until cycle 4, TNS is similar to BTS besides state recording in LIFOs.

In cycle 5, the sorting process becomes different compared with BTS. TNS does not start a new min search iteration from MSB, but from an intermediate tree node stored in LIFOs. The most recently stored CL4 and its corresponding number status are reloaded, where only two numbers (2 and 3) are valid and one number (2) is sorted; hence, we can immediately locate 3 as the next minimum value and label it as sorted. To this point, all the valid numbers of associate with CI4 have been labeled as sorted, therefore CI4 and its corresponding number status can be cleared from the LIFOs. 

In cycle 6, we reload the next available tree node state (CI3) stored in LIFOs. Here we find that only number 6 is valid and has not been sorted. Although DR3 (DR at column 3) doesn't
reach LSB, we design a last number check mechanism (details in \ref{TNS_Hardware}) which enables locating min 6 in cycle 6 without spending another cycle to reach LSB. The state reloading of CI3 in cycle 6 is also cleared.

In cycle 7, we reload the next available tree node state and there are three valid numbers (9, 14, 14) at this point. Since DR2 results include both 0's and 1's, we record this tree node state CI3 and exclude 14 and 14 where bit 1's occur. After number exclusion, there is only number 9 left, and our last data check mechanism allows us to immediately locate it.

In cycle 8, we reload the tree node state CI 3 we just recorded and find all 1's. Therefore this DR3 is skipped and we move to DR4. In cycle 9, DR4 reaches the LSB and we encounter two duplicate numbers. According to BTS process, we will label the first 14 as sorted and then load from LIFO or return to MSB to find the repeated 14. Here we design a repeated data check mechanism (details in \ref{TNS_Hardware}) to allow DR stay in LSB until all repeated numbers are sorted (cycle 10 in this case). It shall be noted that the NE states of the two tree nodes corresponding to repeated numbers are the same in LIFOs in this cycle. If repeated numbers are not last numbers and iterative min/max search needs to continue, an additional cycle is required to reload a sorted tree node state. This additional time cost will become bigger when using the multi-level strategy, which is discussed in detail in \ref{Redundant_SL_TNS}. Overall, TNS only requires 10 cycles to complete sorting while BTS requires 24 cycles for the same dataset, demonstrating significant speedup capability. 

\figurename~\ref{fig:TNStree} shows the diagram of the corresponding 4-bit DR tree and the visited tree nodes for aforementioned sorting example (2, 3, 6, 9, 14, 14). According to the order of recording in LIFOs, the recorded tree nodes are represented as Node1 to Node4. Last number and repeated numbers are also marked in purple and pink, respectively. We can find that the recorded tree nodes are essentially the branching tree node 1's and our TNS is first carried out along the sub-trees on the other branching tree node 0's. Theoretically, as long as there are enough space in LIFOs, we can record all the tree nodes at each branching node, so that we can use the least number of cycles to locate min/max. However, larger LIFOs (bigger parameter $k$) introduce more complex logic and more area/energy cost; on the other hand, the maximum achievable clock frequency will also decrease with large LIFOs. We study the impact of parameter $k$ in Supplementary Section~\ref{Details_Performance}.

\newpage
\section{Timing Comparison of BTS and TNS}\label{Timing_Comp_BTS_TNS}

\begin{figure}[hbt!]
    \centering
    \includegraphics[width = 1\linewidth]{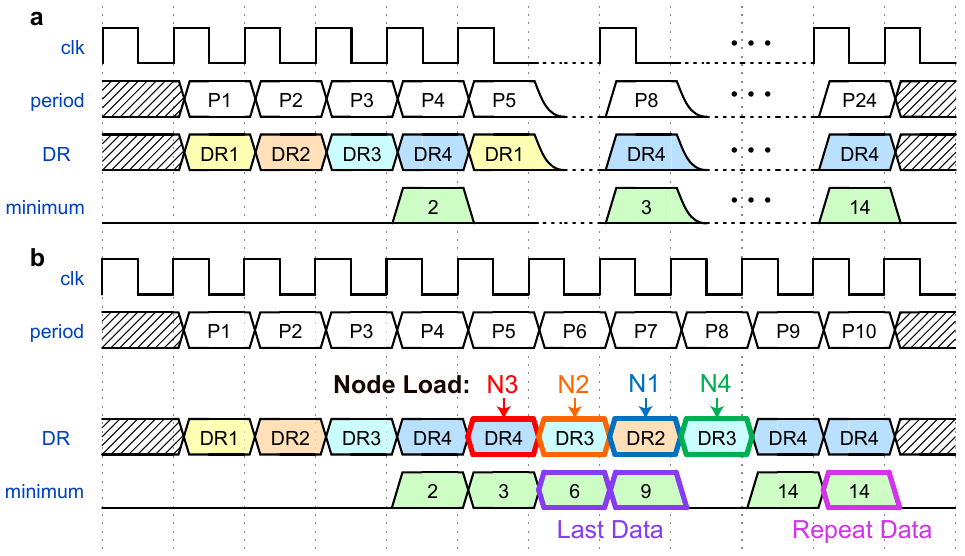}
    \caption{Sorting waveform when using \textbf{a,} BTS and \textbf{b,} TNS.}
    \label{fig:wave}
\end{figure}

The timing schedule of BTS and TNS sorting process for the sample unsigned 4-bit sorting can be seen through the waveform diagram (\figurename~\ref{fig:wave}) more clearly. In TNS, we reload the tree node state in cycle 5, 6, 7, 8 and find the min values in cycle 4, 5, 6, 7, 9, 10. On average, TNS takes less than 2 cycles to sort a number, while BTS takes 4 cycles (\figurename~\ref{fig:wave}) to sort a number. For a deeper DR tree (such as 32-bit numbers), TNS can achieve even better performance because more DRs can be saved while BTS always needs 32 cycles to sort a number.

\newpage
\section{TNS Supporting Variable Data Types}\label{TNS_Data_Types}

\begin{figure}[hbt!]
    \centering
    \includegraphics[width = 1\linewidth]{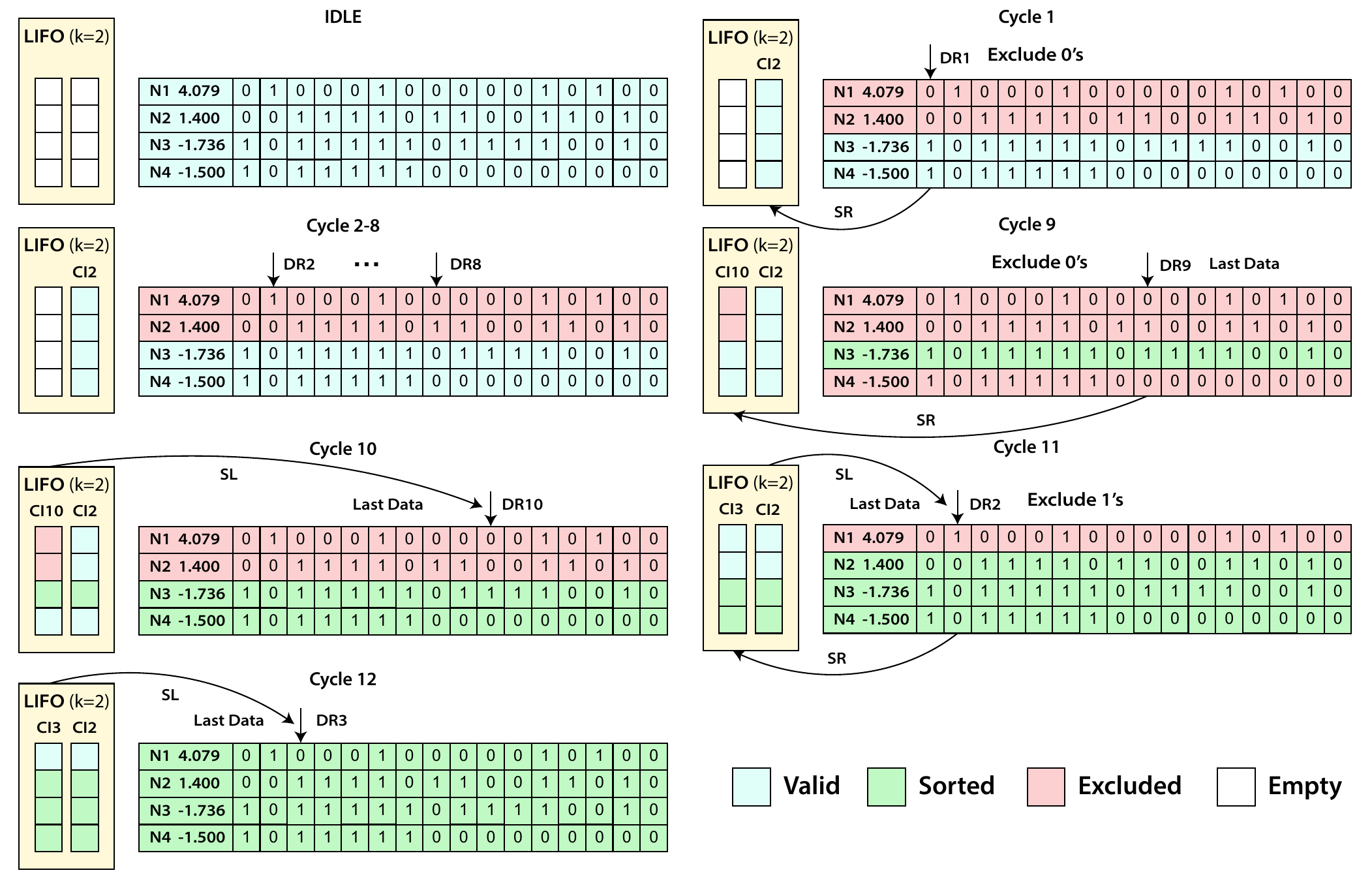}
    \caption{\textbf{Sorting examples with floating-point numbers (This method also works for sign-and-magnitude numbers).}}
    \label{fig:fpexample}
\end{figure}

The simple sorting example introduced earlier are all based on unsigned numbers. In practice, there are many other data types. In \ref{subsubsec2.2.2}, we briefly introduce the methods of extending TNS to sign-and-magnitude, two's complement or floating-point numbers. Here, we provide detailed examples to illustrate how TNS support different data types (\figurename~\ref{fig:fpexample}). Firstly, the IEEE standard for floating-point arithmetic (IEEE 754) is a technical standard for floating-point arithmetic, where floating-point number bits are partitioned into sign bit ($s$), exponent bits ($e$) and fraction bits ($f$). The actual value of a floating-point number can be computed as follows \eqref{fp_compute}:

\begin{equation}
\label{fp_compute}
\text{Floating-Point Value } = (-1)^{s} \times 2^{e-\text{bias}} \times 1.f
\end{equation}

The half-precision floating-point numbers used in the Dijkstra's algorithm (\ref{subsec3.1}) consists of 1 sign bit, 5 exponent bits, and 10 fraction bits, and the bias term is set to 15. Except the sign bit, upper digits are more significant than lower digits. To realize min/max search for floating-point numbers, we need to first separate positive numbers and negative numbers. This is because for positive numbers, the upper digits of 1's correspond to a larger number; but for negative numbers, the upper digits of 1's correspond to a smaller number. Hence, for min search, TNS needs to exclude DR results 0's before all negative numbers are sorted and exclude DR results 1's after all negative numbers are sorted. 

For illustration, we sort four half-precision floating-point numbers in ascending order with TNS $k=2$ as an example (\figurename~\ref{fig:fpexample}). In cycle 1, we exclude 0's other than 1's to search min values in negative numbers, but SR still proceeds as unsigned numbers. In the following cycle 2 to cycle 8, DR results are either both 0's or both 1's; therefore, these DRs are skipped and no action is needed. In cycle 9, we exclude N4 ($-1.500$) and find the first min value N3 which triggers the last number check. Then we reload the most recent tree node state in LIFOs and find the next min value N4 in cycle 10. To this point, we have sorted all negative numbers, therefore we switch to exclude 1's in the following min search iterations. We move back to DR2 and exclude N1 (4.079) in cycle 11. Finally, it takes 12 cycles to sort the four floating-point numbers. This method can also be used for signed-and-magnitude numbers, because their sign bit only represents the positive or negative polarity as in floating-point numbers, not contributing to the magnitude.

\begin{figure}[hbt!]
    \centering
    \includegraphics[width = 1\linewidth]{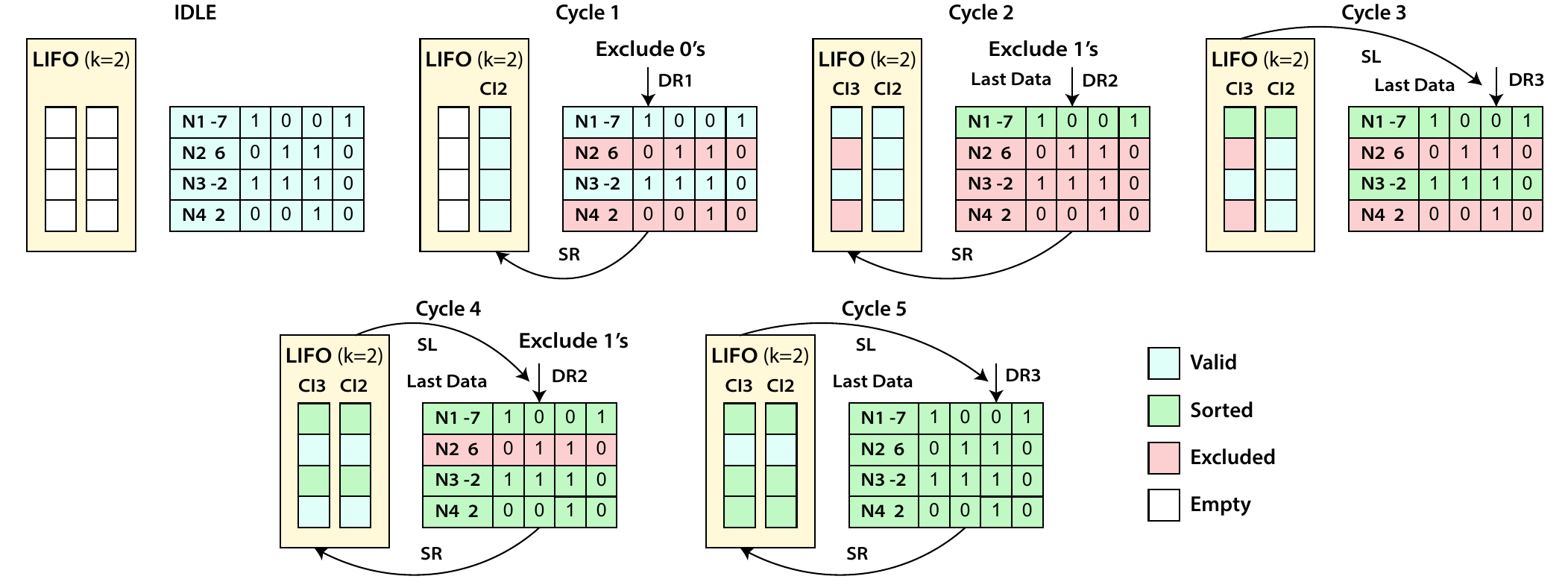}
    \caption{\textbf{Sorting examples with two's complement numbers.}}
    \label{fig:2cexample}
\end{figure}

On the other hand, for signed two's complement $n$-bit numbers with sign bit $s$ and rest of the bits $r$, the sign bit also determines the magnitude of the number as follows \eqref{twos_complement}:

\begin{equation}
\label{twos_complement}
\text{Two's Complement Value } = s \times (-2^{n-1}) + \sum_{i = 0}^{n-2}r_i \times 2^{i}
\end{equation}

\noindent Therefore, bit 1 corresponds to smaller number at the sign bit and we only need to exclude 0's at the sign bit for min search. Here we sort four 4-bit two's complement numbers ascendingly for illustration (\figurename~\ref{fig:2cexample}). In cycle 1, we do DR1 at the sign bit; hence we exclude 0's for min search. Then we go to the next bit and exclude DR results 1's (N3) in cycle 2. At this point, only the last data is left and we find the first min value -7. In cycle 3, we reload the tree node state and locate the last number -2. Furthermore, we go back to DR2 and exclude N2 to locate the next min value N4 in cycle 4. In cycle 5, we finish the sorting of the four numbers.

TNS can also support sorting of other data types following similar methodologies and the associated logic are easy to implement in circuits. Here we implement TNS that can be configured for different data types with slightly different number exclusion logic. Finally, these methodologies are also compatible with three proposed cross-array strategies. 

\newpage
\section{Implementation Details of TNS}\label{TNS_Hardware}

We develop hardware architecture to implement TNS as shown in \figurename~\ref{fig:Fig3}a. The DR currents from the 1T1R array need to be first converted into digital DR signals through SAs or ADCs before sending them to state controller. In the logic module, DR signals enter the \emph{All 0's or 1's} module to determine whether they meet the NE and SR conditions to generate the $ren$ signal. The $ren$ controls the LIFOs to perform state recording and then controls the NE processor to exclude valid numbers. The number status after exclusions are sent to the \emph{Min/Max Check} module to determine if min/max value is located. If so, we output the min/max value and the $len$ to control the operations of the \emph{Load Check} module. The \emph{Load Check} module checks whether the current tree node as LIFO output has been sorted completely; and if not, it sends the \emph{load} signal to reload the state.

For \emph{All 0's or 1's} module, we need to check whether the valid DR results are all 0's or all 1's. For checking all 0's, we need to perform NAND operation using currently valid number status (NE signals) and their corresponding DR signals. Using similar methods, we can also check all 1's. The $ren$ signal can be derived based on all 0's or all 1's check, where $ren$ = 1 indicates that both 0's and 1's exist in the valid DR results. Both NE and SR operations are controlled by the $ren$ signal. The logic of NE is relatively simple, where we only need to set the number status with valid DR results 1's as excluded. For \emph{Load Check} module, we need to read number status of the tree node from the LIFOs and then perform an OR operation (built by NOT and NAND gates), because both the excluded and sorted numbers have number status of logic 0. Therefore, if there are still valid numbers (corresponding to number status of logic 1) for a tree node, this means that the sorting for that tree node and its sub-tree nodes has not been finished and we need to send \emph{load} signal to reload the tree node. 

The \emph{Load Check} module is controlled by the signal $len$ and the generation of $len$ is described in \emph{Min/Max Check} module. We first need to determine whether there is only one number left with number status logic 1 to check if it is the last number. If so, regardless of whether LSB is reached or not, the number left is the next min and then a $len$ signal is sent to determine whether the next reload tree node is ready. Otherwise, if it is not the last number, we will determine whether it has reached LSB. If LSB is not reached, we move on to the next bit for DR; if LSB is reached, we find a new min value, but we still need to determine whether there is only one min value left. In case there are repeated min values, we will stay at LSB in the next cycle until all repetitions are sorted out, after which we send the $len$ signal for \emph{Load Check}.

\newpage
\section{Details on CA-TNS Strategies}\label{Detals_CA_TNS}

To illustrate CA-TNS strategies in detail, we sort four 4-bit unsigned numbers (2, 3, 9, 14) in ascending order as examples. 

\subsection{Multi-Bank Strategy}\label{MB_Strategy}

\begin{figure}[hbt!]
    \centering
    \includegraphics[width = 1\linewidth]{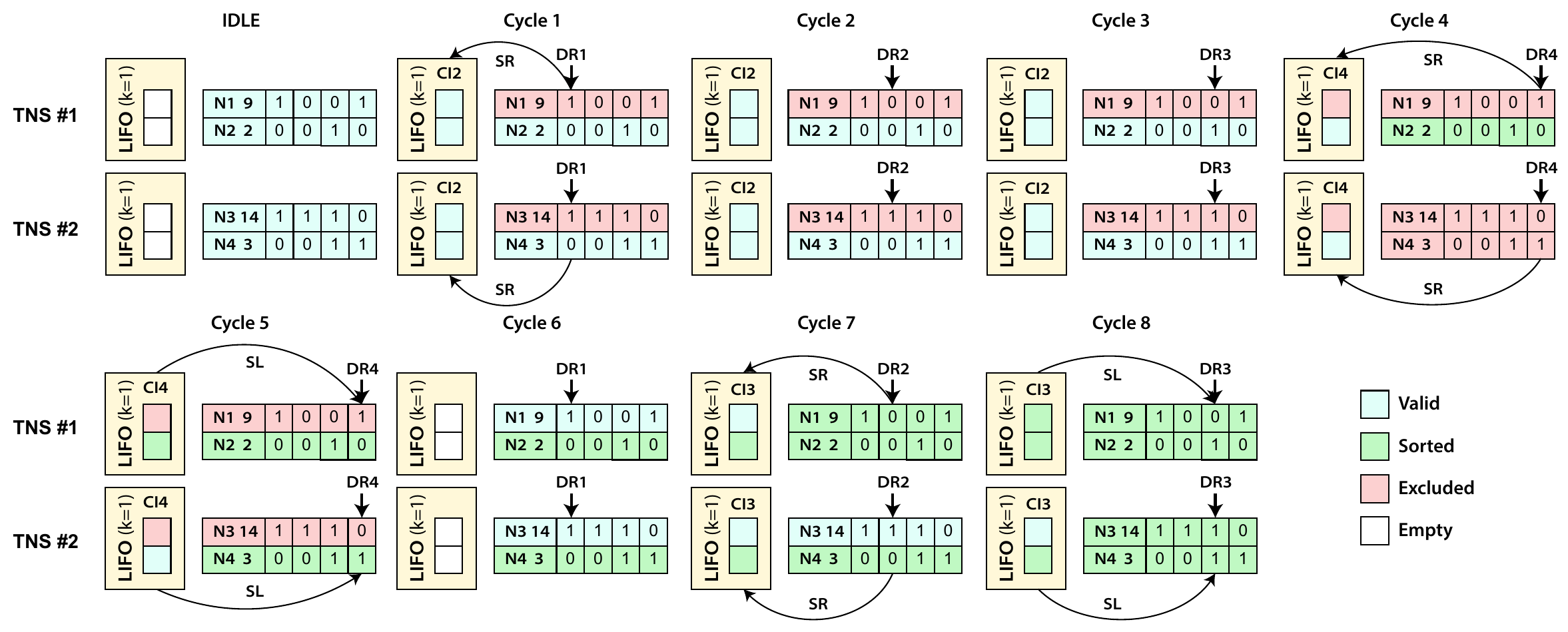}
    \caption{\textbf{Sorting example with multi-bank CA-TNS strategy}.}
    \label{fig:exampleMB}
\end{figure}


The multi-bank strategy aims to solve the scalability problem when data quantity grows up. Here we partition the dataset into different memristor memory arrays based on numbers, where each memristor array has its own periphery circuit and can run as an independent sub-sorter using TNS. For illustration, we partition the example $N=4$ dataset (9, 2, 14, 3) into two ($n = 2$) sub-sorter with $k = 1$: number 9 and 2 in TNS \#1 and number 14 and 3 in TNS \#2 (\figurename~\ref{fig:exampleMB}). 

Note that multi-bank strategy latency to sort $N$ numbers maintains the same as basic TNS, because all operations at the sub-sorters are synchronized and behave like a length-$N$ basic TNS sorter. The all 0's or all 1's check, last number check and repeated number check are carried out across different sub-sorters. In cycle 1, DR results include both 0's and 1's; therefore we record the tree node state and exclude 9 and 14 in sub-sorter \#1 and \#2, respectively. No numbers are exclude in cycle 2 and cycle 3 since the remaining valid numbers have same DR results. In cycle 4, synchronized check across different sub-sorters allow us to exclude number 3, even though bit 0 is located in sub-sorter \#1 while bit 1 is located in another sub-sorter \#2. Meanwhile, this tree node state overwrites the previous record in LIFO because the LIFO size here is only 1. In cycle 5, we reload tree node from LIFOs and locate the min value 3. Since the LIFOs are empty in cycle 6, we restart from MSB and do not exclude any number. In cycle 7, we go to DR2, record the tree node and locate the min value 9. In cycle 8, we reload the tree node state from LIFOs and complete sorting. One can see that the number of DRs with multi-bank strategy remains the same as basic TNS. However, multi-bank strategy suppresses the super linearly of TNS periphery with N, improving the achievable clock frequency and leading to slightly shorter latency than basic TNS.

\newpage
\subsection{Bit-Slice Strategy}\label{BS_Strategy}

\begin{figure}[hbt!]
    \centering
    \includegraphics[width = 1\linewidth]{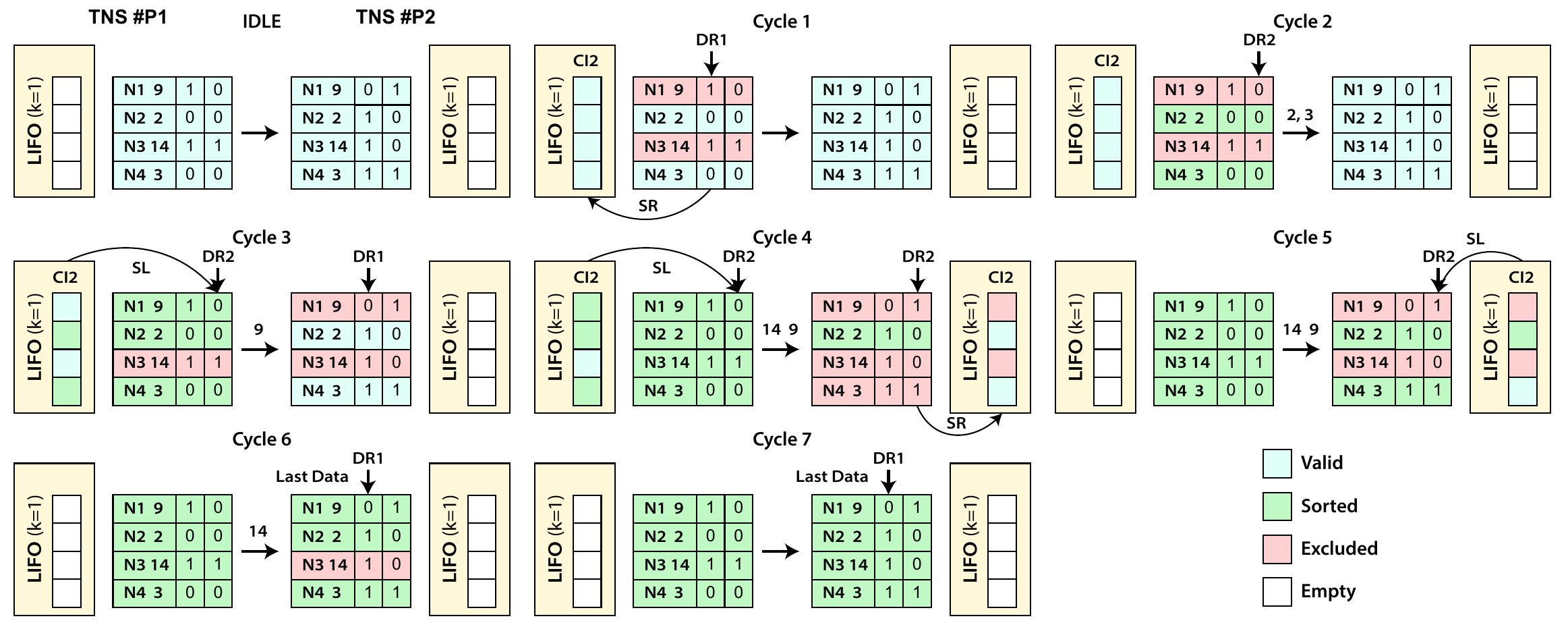}
    \caption{\textbf{Sorting example with bit-slice CA-TNS strategy}.}
    \label{fig:exampleBS}
\end{figure}

Using the bit-slice strategy, the same dataset is sorted in parallel (\figurename~\ref{fig:exampleBS}). The dataset is partitioned into two parts according to digit positions as shown in the example: we store the upper 2 bits of all data in TNS \#P1 and the lower 2 bits in TNS \#P2. Similar to MB strategy, the two TNS sub-sorter can run independently with their own periphery. If TNS \#P1 finds the unique min value locally, the current min search iteration stops and moves to the next iteration; Otherwise, if the local min value at TNS \#P1 has multiple repetitions, we transfer number status to TNS \#P2 through a FIFO and continue the min search in TNS \#P1. 

In the example, TNS \#P1 stores the upper 2 bits of the four numbers (2, 3, 9, 14), while TNS \#P2 stores the lower 2 bits. In cycle 1, we start DR1 and carry out SR and NE operations in TNS \#P1, while TNS \#P2 is still in IDLE state. We find the two local min values N2 and N4 which share the same upper 2 bits in cycle 2; hence, there number status are transferred to TNS \#P2 for further search. In cycle 3, TNS \#P1 reloads the tree node state and find the min value 9, while TNS \#P2 starts DR1 and does not exclude any number. One may note that, from cycle 3, the two sub-sorters start to run in a pipelined manner. In cycle 4, TNS \#P1 finds the last min value (14) and finishes its work. TNS \#P2 excludes 3 and finds the min value 2. In cycle 5, TNS \#P1 goes back to IDLE state and we reload the tree node state in TNS \#P2 and find the min value 3. In the following two cycles (cycle 6 and cycle 7), TNS \#P2 outputs 9 and 14 and then completes the sorting process. Compared to multi-bank strategy or basic TNS, the bit-slice strategy only requires 7 cycles to complete the same sorting task, because the pipelined sorting introduced by bit-slice strategy introduces higher level of parallelism.

\newpage
\subsection{Multi-Level Strategy}\label{ML_Strategy}

\begin{figure}[hbt!]
    \centering
    \includegraphics[width = 1\linewidth]{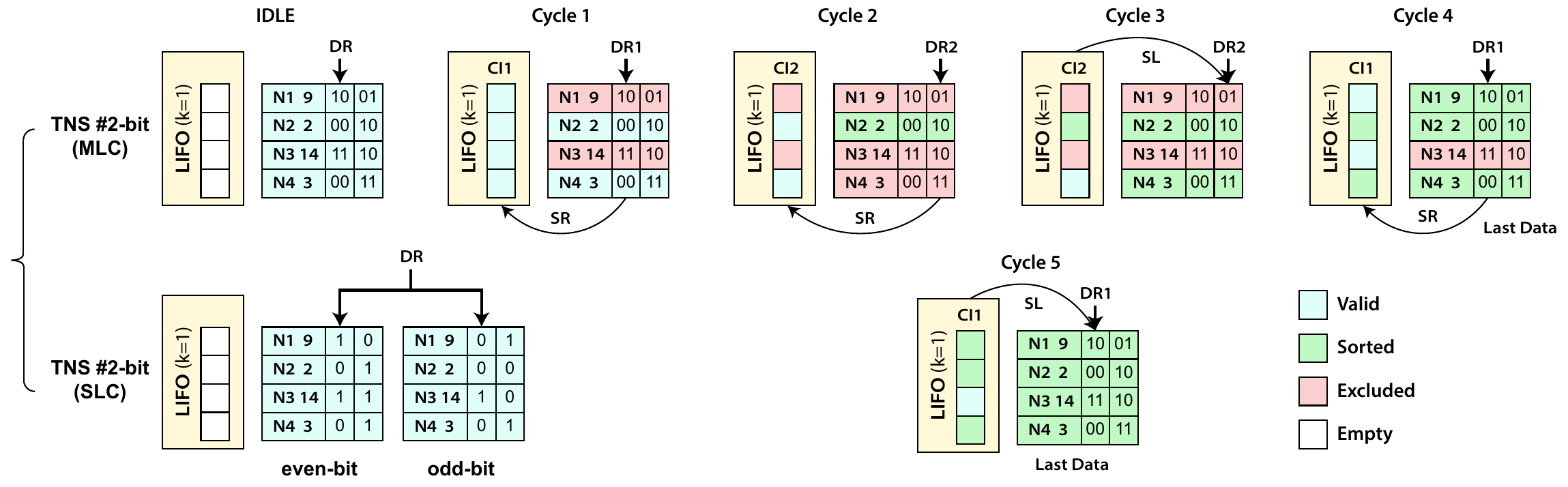}
    \caption{\textbf{Sorting example with multi-level CA-TNS strategy}.}
    \label{fig:exampleML}
\end{figure}

For multi-level strategy, we introduce two array mapping schemes (\ref{subsubsec3}) to retrieve more than one bits for each DR. The first scheme is to use multi-level memristor devices and the second one is to read multiple single-level devices simultaneously (\figurename~\ref{fig:exampleML}). Taking ML-$2$-bit for illustration where each 4-bit number can be stored using two memristor devices. In cycle 1, we start DR1 and DR results have 3 different values, 10's, 00's and 11's. Unlike NE operations in basic TNS, here we need to exclude all the larger DR results and leave only the smallest one. So we exclude 9 and 14 and record this tree node state. 

Note that when using the multi-level strategy, we record the current column index (CI) instead of next CI. This is because the DR tree is no longer a binary tree but a quad tree in multi-level cases. After we finish a sub-tree at a branching tree node, there are multiple other sub-trees left and we can not skip this branching tree node directly. Using the same NE logic, we exclude 3 and find the min value 2 in cycle 2. In cycle 3, we reload the tree node state and find the min value 3. Now the LIFOs are empty and we start DR from MSB and exclude 14 to find the next min value 9. In cycle 5, we finish the entire sorting process. Due to the fact that we can retrieve twice as much information per DR as before, the sorting cycles required are also greatly reduced to only 5 for the same dataset. However, the required NE logic is much more complex to handle multi-bit DR results, leading to degraded clock frequency. Here we further introduce pseudo multi-level processing by using multiple single-level memristors. In this work, we demonstrate ML strategy up to 8 levels. Further extending the NE logic and the ADC resolutions can also support devices with more levels, but may degrade sorting accuracy due to overlapped conductance states.

\newpage
\section{Implementation Details of CA-TNS}\label{CA_TNS_Hardware}

To implement the three CA-TNS strategies introduced in \ref{subsec2}, we extend circuit design on top of basic TNS implementations (\figurename~\ref{fig:Fig3}c). For multi-bank strategy, we need a cross-array processor configured as multi-bank mode to synchronize necessary operations across different sub-sorters (\figurename~\ref{fig:Fig3}c). Suppose there are $n$ TNS sub-sorters, their \emph{not all 0's}, \emph{not all 1's}, and \emph{load} signals need to be sent to cross-array processor for synchronization. The above three signals of sub-sorters go through an OR logic (use NOT and NAND gates to implement) to obtain three synchronized signals. The OR logic makes sure that if one sub-sorter needs to do an operation, all the others can follow. These three synchronized signals are then sent back to sub-sorters to replace the original control signals. An output controller is also needed to handle sorting output from different sub-sorters. 

For bit-slice strategy with $n$ partitions, we need $n$-1 cross-array processor configured as bit-slice mode to store the intermediate number status from predecessor sub-sorter (\figurename~\ref{fig:Fig3}c). Each time TNS \#P1 finds min values, they are stored in NE FIFOs. The min values found in TNS \#P1 generally has many repetitions, because some numbers may have the same upper bits and cannot be distinguished by TNS \#P1. The number status stored in NE FIFOs are used to initialize the TNS \#P2. After that, TNS \#P2 starts to sort the remaining valid numbers from TNS \#P1 and outputs number status to next sub-sorter if needed. One may note that this implementation relies on the sizes of NE FIFOs where we need to ensure that the FIFOs are large enough for sub-sorters to work in a pipelined manner. In this work, we instantiate FIFOs with enough space to support bit-slice strategy and study the area and energy efficiency cost in Supplementary Section~\ref{Details_Performance}.

For multi-level strategy, we need to change the digit processor and state controller (\figurename~\ref{fig:Fig3}a) to support multi-bit DR results. ML-$n$-bit introduces $2^{n}$ levels at each memristor device and $n$-bit ADCs are required to convert analog DR signals into digital values in digit processor. In state controller, we extend the logic of number exclusion. Firstly, we perform \emph{all 0's or 1's} check on each bit of DR results and obtain the bit-wise enable signals from $en[n-1]$ to $en[0]$. Secondly, we carry out $n$-bit number exclusion from digit $n-1$ to digit $0$ of DR results.Note that as $n$ increases, although we retrieve more information from each DR, the processing logic also becomes more complex, resulting in degraded area and clock frequency. We evaluate the ML strategy performance in Supplementary Section~\ref{Details_Performance}.


\newpage
\section{System Design and Experimental Setup}\label{PCB}

\begin{figure}[hbt!]
    \centering
    \includegraphics[width = 0.8\linewidth]{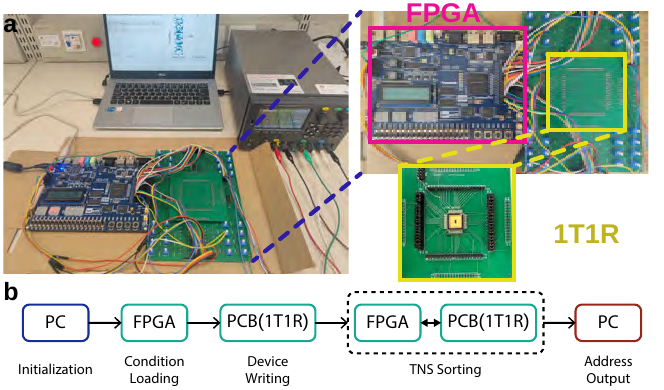}
    \caption{\textbf{a,} Photos of our MSIM system. \textbf{b,} The workflow and interaction process of our MSIM system.}
    \label{fig:system}
\end{figure}

We build a memristor-based hardware and software co-designed SIM system to implement proposed TNS/CA-TNS schemes (\figurename~\ref{fig:system}). We build periphery read and write circuits for our 1T1R chip on the PCB that can interact with FPGAs. FPGAs control the read and write of our memristor chip and process the signals from PCB to complete sorting process. We use a PC to initialize sorting configurations to FPGAs including sorting dataset, encoding method (data types) and sorting strategy. Here PC controls FPGA for device writing. FPGAs control subsequent TNS/CA-TNS processing for sorting results, which are sent back to PC and used in different real-world applications.

\begin{figure}[hbt!]
    \centering
    \includegraphics[width = 0.8\linewidth]{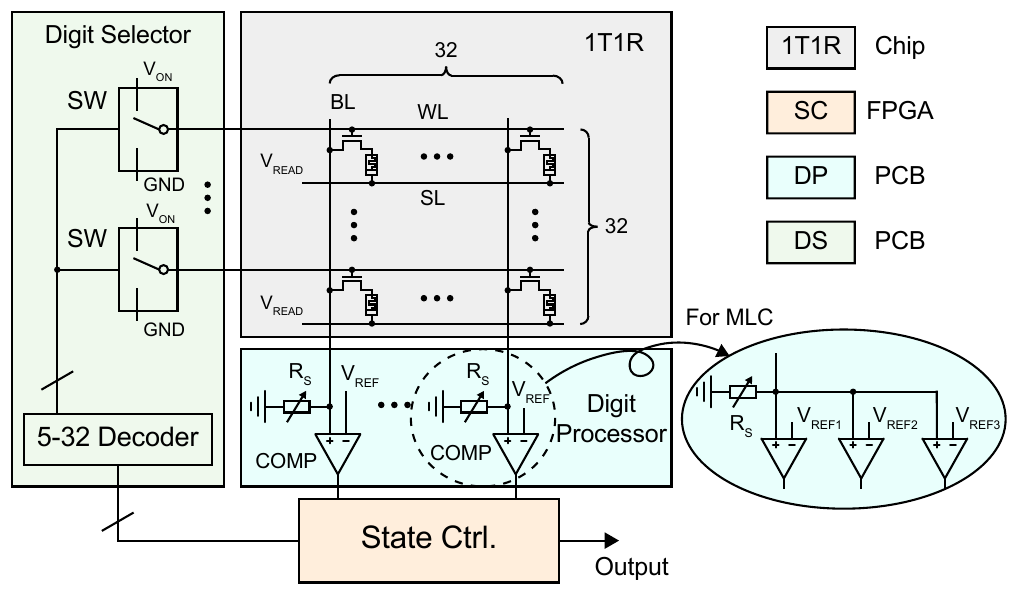}
    \caption{\textbf{The specific implementation methods of the four modules of TNS}.}
    \label{fig:PCB}
\end{figure}

\vspace{2.0em}
\begin{minipage}{\textwidth}
\begin{minipage}[t]{0.5\textwidth}
\makeatletter\def\@captype{table}
\caption{All components and models used on the PCB.}
\includegraphics[width=\textwidth]{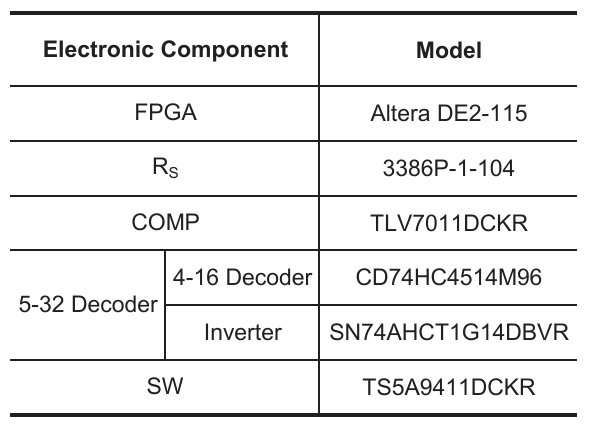}
\label{tab:model}
\end{minipage}
\begin{minipage}[t]{0.4\textwidth}
\makeatletter\def\@captype{table}
\caption{Selection of various parameters in the experiment.}
\includegraphics[width=\textwidth]{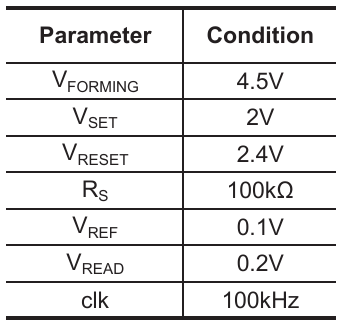}
\label{tab:condition}
\end{minipage}
\end{minipage}

Our TNS/CA-TNS hardware consists of four modules, 1T1R memristor chip, digit processor, digit selector and state controller (\figurename~\ref{fig:PCB}). Using our 32$\times$32 1T1R chip, we design corresponding read and write circuits on PCB. The detailed implementation architecture is shown in \figurename~\ref{fig:PCB}. We use sampling resistors and comparators (COMP) to convert the read currents from our memristor array into digital voltage signals (in digit processor). For single-level devices, we build a simple sense amplifier using a sampling resistor and a comparator. For multi-level devices, comparator arrays are required to distinguish multiple levels. The state controller is implemented in FPGAs which output the WL addresses for the next DR and pass them to a 5-32 decoder built from two 4-16 decoders and an inverter. Each WL is controlled by an analog switch (SW) to turn on and off. The 5-32 decoder determines which row is read by controlling 32 analog switches (in digit selector). The models and parameters of electronic components we use on PCB are shown in \tablename~\ref{tab:model} and \tablename~\ref{tab:condition}. Note that although our experiment is conducted at 100kHz, it is actually limited by the electronic components and parasitic resistors/capacitors on the PCB. TNS/CA-TNS can run in a higher clock frequency if we integrate them into a full functional chip.

\newpage
\section{Detailed Performance Evaluations}\label{Details_Performance}

\subsection{Sorting Speed Evaluations}\label{Speed_Eval}

We evaluate sorting speed using our end-to-end memristor-based TNS/CA-TNS SIM system for low ($W$ = 8 bits) and high ($W$ = 32 bits) precision datasets. Here we carry out experiments across five datasets including three classical statistically-distributed datasets (random, normal, and clustered) and two widely-used sorting benchmark datasets (Kruskal's and MapReduce). 

\subsubsection{Low-Precision Experiments}\label{LP_speed}

\begin{figure}[hbt!]
    \centering
    \includegraphics[width = 1\linewidth]{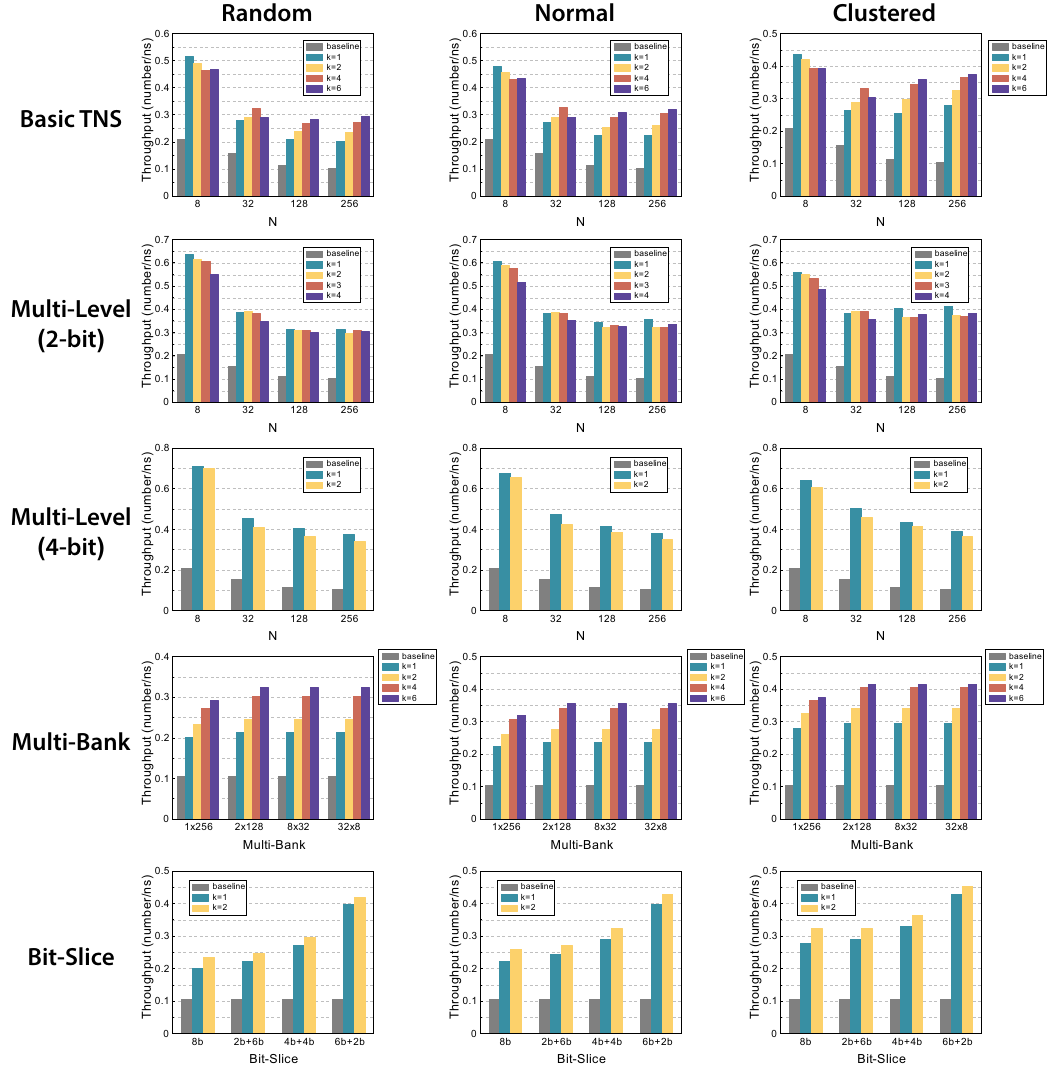}
    \caption{\textbf{Speed evaluation of sorting on 8-bit data using CA-TNS in 3 datasets (Random, Normal and Cluster). We study the variation of sorting speed with the number of data $N$ and $k$ with the basic TNS and Multi-Level strategy. For Multi-Bank strategy, we study the speed to sort a total of 256 numbers for different division cases. We also investigate the effect of different width divisions on the sorting speed for sorting a total of 256 numbers with Bit-Slice strategy.}}
    \label{fig:8bit_speed}
\end{figure}

We first evaluate the sorting speed using basic TNS and investigate the effect of data quantity ($N$) and TNS parameter $k$ (\figurename~\ref{fig:8bit_speed}). From the experimental results, the sorting speed increases with $k$ in general. With $k$ increases, more tree nodes are recorded so that it is more likely to start min/max search iteration from an intermediate tree nodes closer to the min/max values. When $N$ is small, the depth-$W$ DR tree has many nodes which are not easy to be reused, so it takes longer to search all the branches. But when $N$ is large enough to hold most of the DR tree, the gain from recording tree nodes becomes more significant. However, increasing $N$ or $k$ degrades the operating frequency of TNS hardware, which may in turn lower the sorting speed. 

For multi-level strategy, we also evaluate the sorting speed for ML-$2$-bit and ML-$4$-bit cases. As the cell level increases, although more information can be retrieved per DR and fewer DRs are required for sorting, the periphery circuitry are much more complex while the achievable operating frequency decreases. Taking the lower clock frequency factor into consideration, the sorting speed with ML strategy is still better than the basic TNS. However, the sorting speed with ML strategy decreases more significantly with increasing $k$. This is because the speed increase from increasing $k$ does not overcome the reduction in operating frequency and there are more redundant DRs in ML strategy (see \ref{Redundant_SL_TNS} for more details).

For multi-bank strategy, we evaluate four different partition choices that implement the sorting of 256 8-bit numbers (1 $\times$ length-256 sorter, 2 $\times$ length-128 sorters, 8 $\times$ length-32 sorters and 32 $\times$ length-8 sorters). Although the multi-bank strategy takes the same number of DRs as the basic TNS, differences in the periphery circuits result in different operating frequencies, where 32 $\times$ length-8 sorters result in fastest sorting speed but the speedup also gets saturated when reaching 32 sub-sorters.

For bit-slice strategy, we partition the dataset into two parts (upper digits and lower digits) to evaluate the sorting speed (1 $\times$ 8-bit sorter, 1 $\times$ 2-bit sorter + 1 $\times$ 6-bit sorter, 1 $\times$ 4-bit sorter + 1 $\times$ 4-bit sorter and 1 $\times$ 6-bit sorter + 1 $\times$ 2-bit sorter). We find that increasing $k$ has smaller effects on the speedup for bit-slice strategy. Fast sorting speed can be achieved with a low $k$. However, the partition methodology for upper and lower bits has a very strong effect on the sorting speed. Across experimental datasets, 1 $\times$ 6-bit sorter + 1 $\times$ 2-bit sorter achieves the highest speedup performance, reaching almost twice as much as the basic TNS speed.

\newpage
\subsubsection{High-Precision Experiments}\label{HP_speed}

\begin{figure}[hbt!]
    \centering
    \includegraphics[width = 1\linewidth]{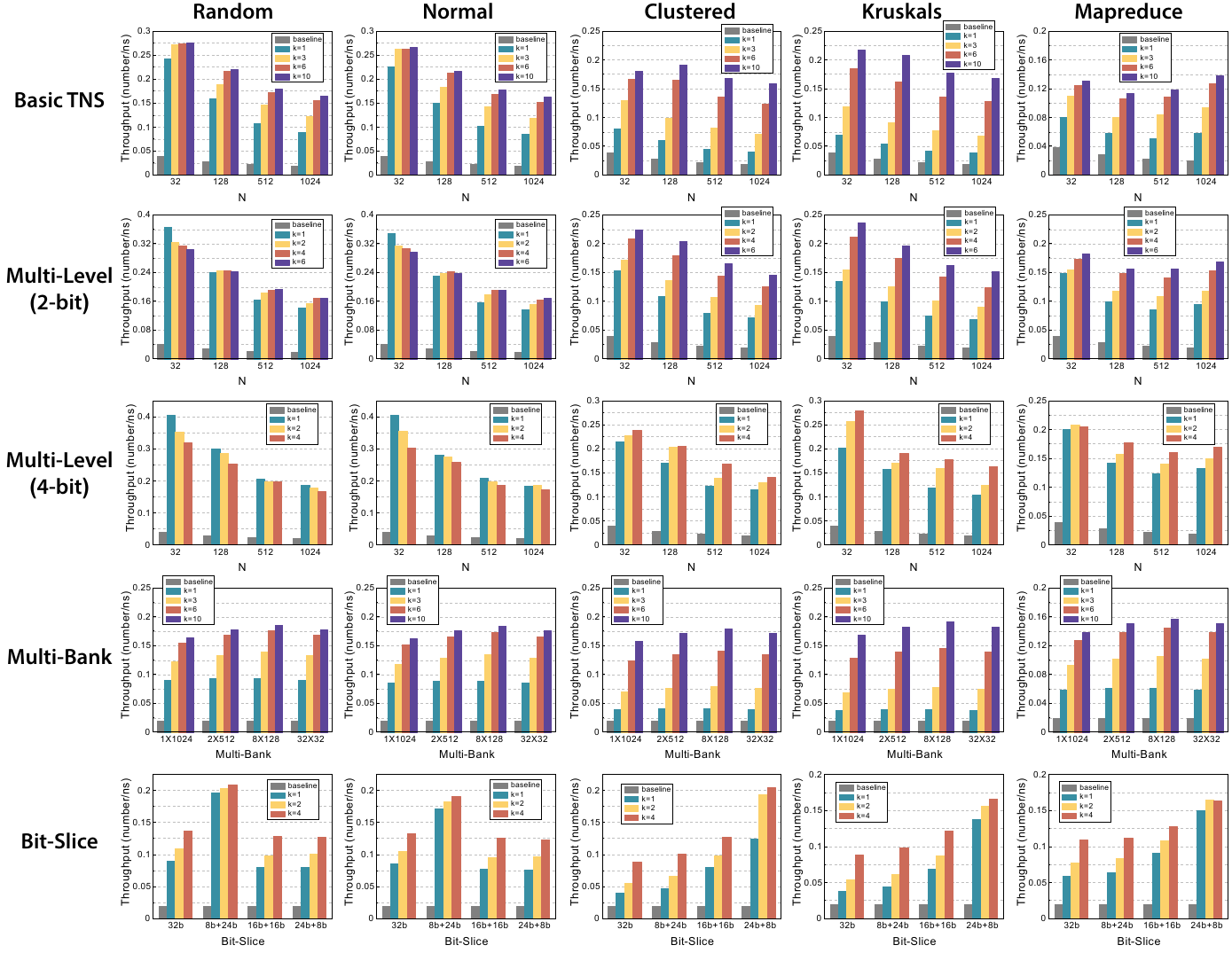}
    \caption{\textbf{Speed evaluation of sorting on 32-bit data using CA-TNS in 5 datasets (Random, Normal, Cluster, Kruskals and Mapreduce). We study the variation of sorting speed with the number of data $N$ and $k$ with the basic TNS and Multi-Level strategy. For Multi-Bank strategy, we study the speed to sort a total of 1024 numbers for different division cases. We also investigate the effect of different width divisions on the sorting speed for sorting a total of 1024 numbers with Bit-Slice strategy}}
    \label{fig:32bit_speed}
\end{figure}

Here we report the sorting speed of high-precision 32-bit numbers across five datasets (\figurename~\ref{fig:32bit_speed}). For basic TNS, the speedup trends with high-precision numbers are similar to those from low precision experiments, except that the speedup magnitudes are more significant. On the other hand, the performance improvements brought by increasing $k$ are also greater. We also find that TNS performs better on random and normal datasets; but for the other three datasets, a higher $k$ is required to achieve noticeably better results than BTS baseline because the other three datasets are more clustered and there are more DR tree nodes need to be recorded around cluster centers.

For ML-$2$-bit and ML-$4$-bit multi-level strategy, we observe that the sorting speed becomes slower as $k$ increases, especially in the random and normal datasets. This is not just due to the decrease in operating frequency, but also due to a redundant recording and reloading phenomenon which discussed in detail in \ref{Redundant_SL_TNS}.

Similarly, we test 1 $\times$ length-1024 sorter, 2 $\times$ length-512 sorters, 8 $\times$ length-128 sorters and 32 $\times$ length-32 sorters to evaluate multi-bank strategy. The results are consistent with low precision and number-based partitioning enables faster sorting speeds than basic TNS. For bit-slice strategy we also divide the 32-bit dataset into upper digits and lower digits. We find that sorting performance is strongly correlated with the partition bit location and the best partition bit location varies across different datasets. For random and normal datasets, 1 $\times$ 8-bit sorter + 1 $\times$ 24-bit sorter achieves a maximum speedup of nearly doubled speed compared to basic TNS. But for the other 3 datasets, the division of 24-bit + 8-bit shows better performance. But improper division has little negative impact on sorting performance.

\newpage
\subsection{Area and Energy Efficiency Evaluations}\label{Area_Power_Eval}

In this article, we also perform detailed area and energy efficiency evaluations for TNS/CA-TNS with low ($W$ = 8 bits) and high ($W$ = 32 bits) precision numbers. Sorting datasets with higher data precision requires more area and cost more energy. Area and energy evaluations are based on in-lab measurement results and Synopsys Design Compiler. Energy efficiency evaluations are based on random dataset. Area efficiency is defined as $throughput/area$ and energy efficiency is defined as $throughput/power$.


\subsubsection{Low-Precision Experiments}\label{LP_area_power}

\begin{figure}[hbt!]
    \centering
    \includegraphics[width = 1\linewidth]{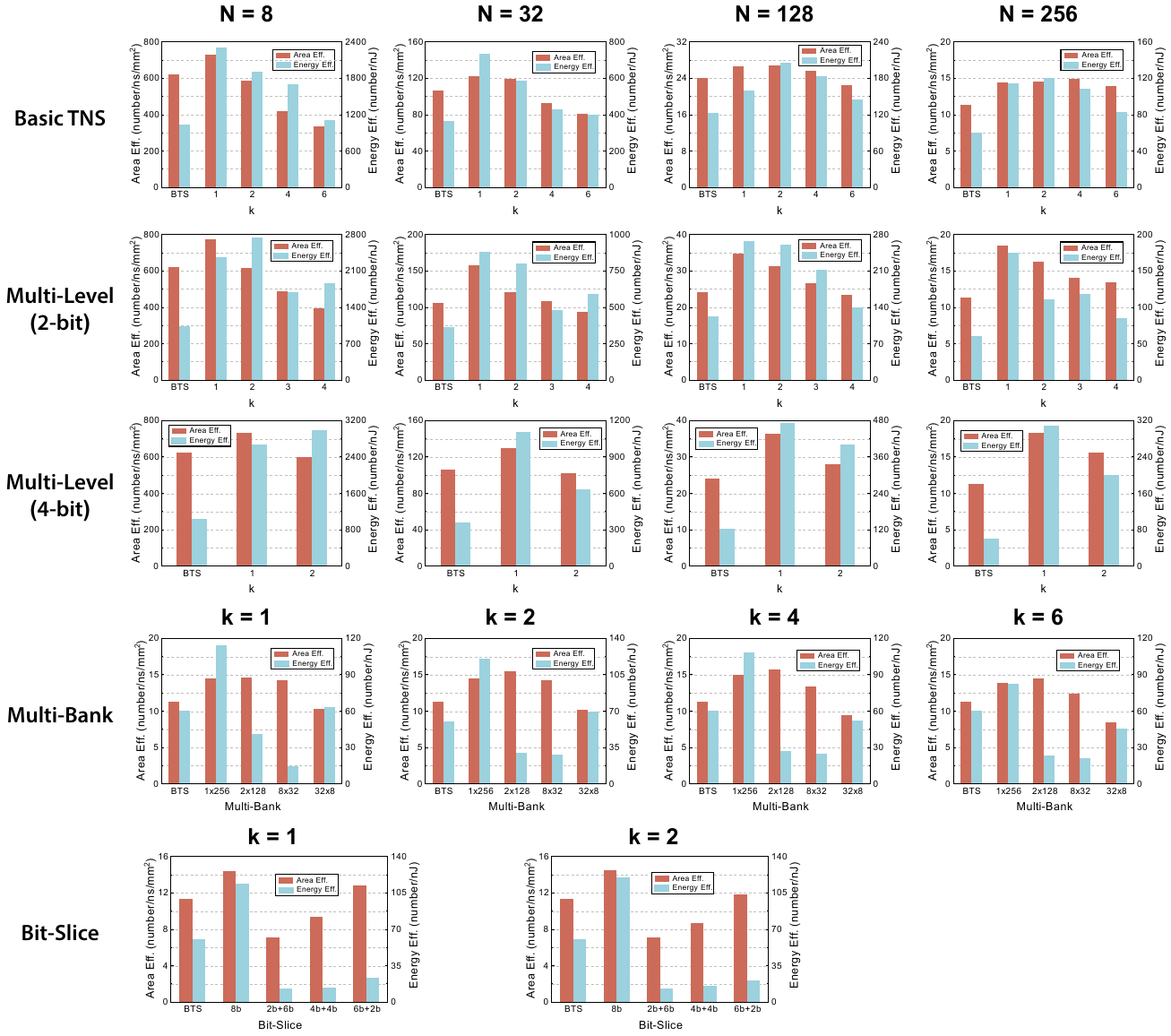}
    \caption{\textbf{Area efficiency and energy efficiency evaluations of sorting datasets with 8-bit precision using TNS/CA-TNS.}}
    \label{fig:8bit_area}
\end{figure}

We first evaluate the area and energy efficiency of basic TNS and investigate the effects of data quantity ($N$) and $k$ (\figurename~\ref{fig:8bit_area}). From the experimental results, the area increases with $k$ increases as expected. This leads to larger area efficiency at lower $k$. Similarly the energy efficiency also goes down as $k$ increases. Meanwhile, both area efficiency and energy efficiency decrease as $N$ increases because the sorting speed of TNS remain approximately the same while area and power consumption go up.

For multi-level strategy, as the cell level increases, although the sorting speed is much faster, the area of the digit processor and state controller are also growing, resulting in a slight increase in area efficiency. However, energy efficiency increases significantly, especially for ML-$4$-bit cases due to the fact that the DR conversions of multi-bits does not consume significant higher energy compared to those in single-level cases.

For multi-bank strategy, we evaluate four different partition choices that implement the sorting of 256 8-bit numbers (1 $\times$ length-256 sorter, 2 $\times$ length-128 sorters, 8 $\times$ length-32 sorters and 32 $\times$ length-8 sorters). We observe that 2 $\times$ length-128 sorters deliver the highest area efficiency across different $k$. As the number of sub-sorters increases, the area efficiency goes down because the area of cross-array processor becomes larger. For energy efficiency, multi-bank strategy leads to worse energy efficiency than the basic TNS (1 $\times$ length-256 sorter) because of the energy consumption of the cross-array processor.

For bit-slice strategy, we partition the dataset into two parts (upper digits and lower digits) to evaluate the sorting speed (1 $\times$ 32-bit sorter, 1 $\times$ 8-bit sorter + 1 $\times$ 24-bit sorter, 1 $\times$ 16-bit sorter + 1 $\times$ 16-bit sorter and 1 $\times$ 24-bit sorter + 1 $\times$ 8-bit sorter).  Since the area and power of TNS sorters do not vary much with data width $W$, both area efficiency and energy efficiency with BS strategy are nearly doubled compared with basic TNS. The differences between different partition choices in \figurename~\ref{fig:8bit_area} is mainly due to the varying speedup when sorting random dataset.

\newpage
\subsubsection{High-Precision Experiments}\label{HP_area_power}

\begin{figure}[hbt!]
    \centering
    \includegraphics[width = 1\linewidth]{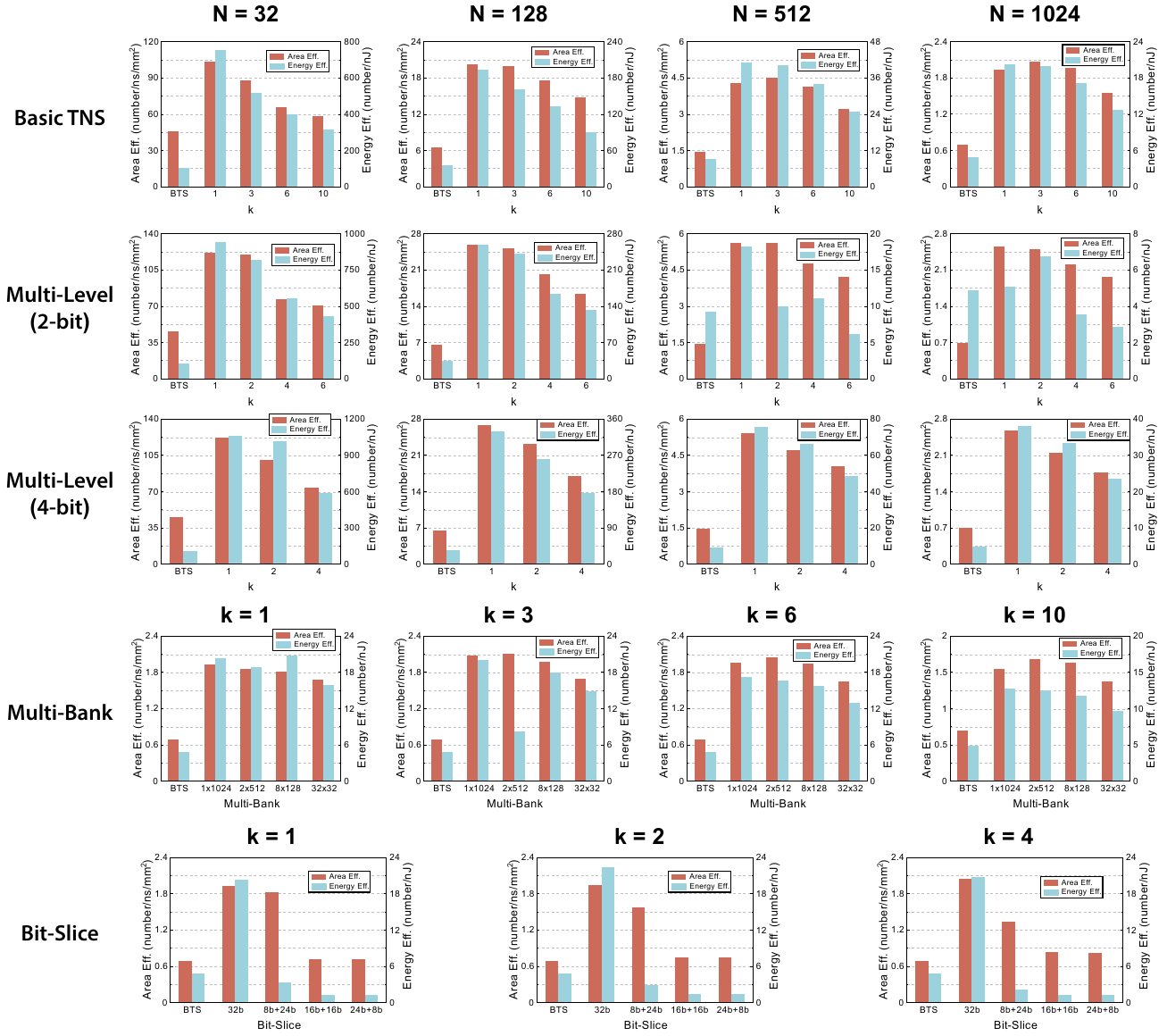}
    \caption{\textbf{Area efficiency and energy efficiency evaluation of sorting on 32-bit data using TNS/CA-TNS.}}
    \label{fig:32bit_area}
\end{figure}

Here we investigate the area and energy efficiency of sorting high-precision datasets using TNS/CA-TNS on random datasets (\figurename~\ref{fig:32bit_area}). For basic TNS, the trends of area and energy efficiency with increasing $k$ are the same as in low precision datasets, but the area and energy efficiency enhancements are much better when comparing to basic TNS. Due to the greatly improved sorting speed, the area and energy efficiency of basic TNS with high-precision datasets are far better than BTS across different $N$.

For multi-level strategy, we observe that the advantages of area efficiency and energy efficiency are more significant especially for ML-$4$-bit cells. One may also observe that for ML-$2$-bit cells, large $N$ degrades the energy efficiency compared to the basic TNS. This is mainly because the periphery processing consumes more energy as $N$ increases but at the speedup increase is not enough.

For multi-bank strategy, we evaluate four different partition choices that implement the sorting of 1024 32-bit numbers (1 $\times$ length-1024 sorter, 2 $\times$ length-512 sorters, 8 $\times$ length-128 sorters and 32 $\times$ length-32 sorters). From the experimental results, the area and energy efficiency improvements over basic TNS are smaller than those in low-precision datasets. This is because the cross-array processor for high-precision datasets are more costly in terms of area and energy consumption.

For bit-slice strategy, we partition the dataset into two parts (upper digits and lower digits) to evaluate the sorting speed (1 $\times$ 32-bit sorter, 1 $\times$ 8-bit sorter + 1 $\times$ 24-bit sorter, 1 $\times$ 16-bit sorter + 1 $\times$ 16-bit sorter and 1 $\times$ 24-bit sorter + 1 $\times$ 8-bit sorter). The experimental results are similar to those for low-precision datasets. The area efficiency nearly triples comped with BTS for 1 $\times$ 8-bit sorter + 1 $\times$ 24-bit sorter, but decreases for 1 $\times$ 16-bit sorter + 1 $\times$ 16-bit sorter and 1 $\times$ 24-bit sorter + 1 $\times$ 8-bit sorter. On the other hand, the energy efficiency when using BL strategy are very low compared to basic TNS due to the energy cost of multiple FIFOs between sub-sorters. The biggest advantage of BS strategy is that it delivers the fastest sorting speed.

\newpage
\section{Tree Node State Recording/Reloading}\label{Redundant_SL_TNS}

In previous Section~\ref{subsec1}, we introduce a parameter $k$ for tree node state recording and reloading. For basic TNS and MB/BS CA-TNS strategies, the number of DRs can usually be reduced with increasing $k$, but the area and energy efficiency generally degrade. This is because when $k$ increases, we have larger LIFOs to record more tree node states which enables more DR saving for faster sorting speed. This is also consistent with what we observe in the performance evaluations (Supplementary Section~\ref{Details_Performance}). 

However, for ML CA-TNS strategy in \ref{Speed_Eval}, the number of DRs spent on sorting does not monotonically decrease with LIFO size $k$. When we adopt multi-level strategy, we observe that as $k$ increases, sometimes the number of sorting cycles increases too especially when using more conductance states. The reason is that for multi-level memristor cells, it is easier to record duplicate states during SR process, which introduces more latency to handle these duplicate states. This is essentially because the DR tree has more than two branches in ML strategy and we are not able to identify which sub-tree to record at a tree node. When a branching sub-tree of a tree node is sorted, these branches become duplicated. This scenario has little impact when using binary cells as the DR tree has only two branches, but the impacts become more significant when using ML devices. 

\begin{figure}[hbt!]
    \centering
    \includegraphics[width = 1\linewidth]{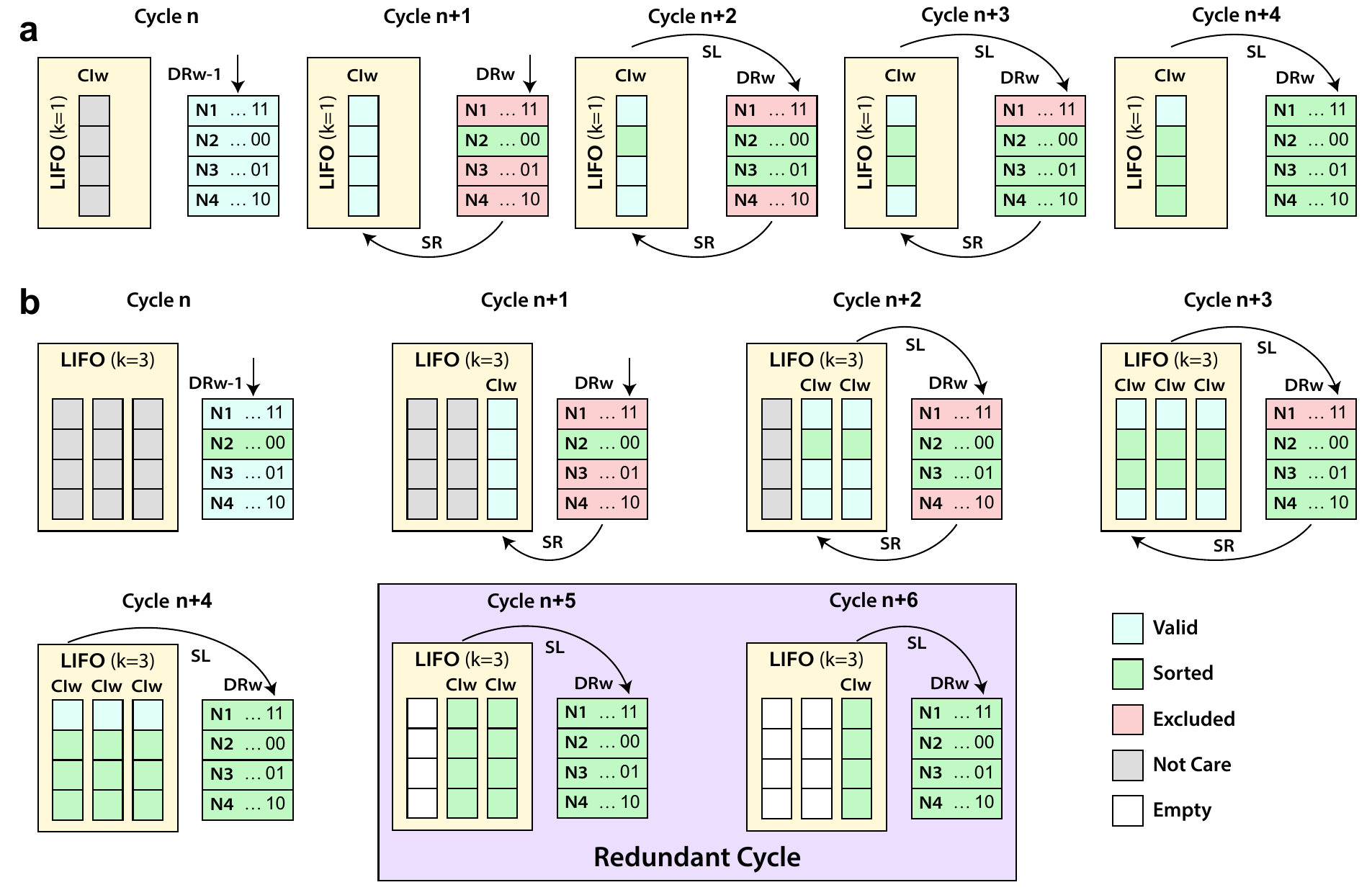}
    \caption{\textbf{An example when the LIFO size ($k$) increases but the sorting speed decreases when using multi-level strategy}. \textbf{a,} $k=1$. \textbf{b,} $k=3$.}
    \label{fig:multibit}
\end{figure}

For illustration purpose, we present an example in \figurename~\ref{fig:multibit}. Taking ML-$2$-bit as an example, suppose start from a sub-tree of a tree node in cycle $n$. If $k=1$, in cycle $n+1$, we do DR$w$, recording this tree node state and excluding $N1$, $N3$ and $N4$. After that, we reload the tree node state, record the new tree state and exclude $N1$ and $N4$ to locate the next min value $N3$ in cycle $n+2$. We perform similar operations to located $N4$ and $N1$ in cycles $n+3$ and $n+4$, respectively. Until this point, we spend $4$ cycles to sort the entire sub-tree, and the LIFOs are empty. 

On the other hand, if we increase $k$ to $k=3$, ML TNS sorting process are similar in the first four cycles as $k=1$, but the only difference is that the newly stored tree node states do not need to be replaced as the LIFO size $k$ is big enough (\figurename~\ref{fig:multibit}b). To this point, there are 3 duplicate states in LIFOs after finishing the sub-tree. Therefore we need two redundant cycles (cycle $n+5$ and $n+6$) to clear the LIFOs for subsequent min/max search iterations.

\begin{figure}[hbt!]
    \centering
    \includegraphics[width = 1\linewidth]{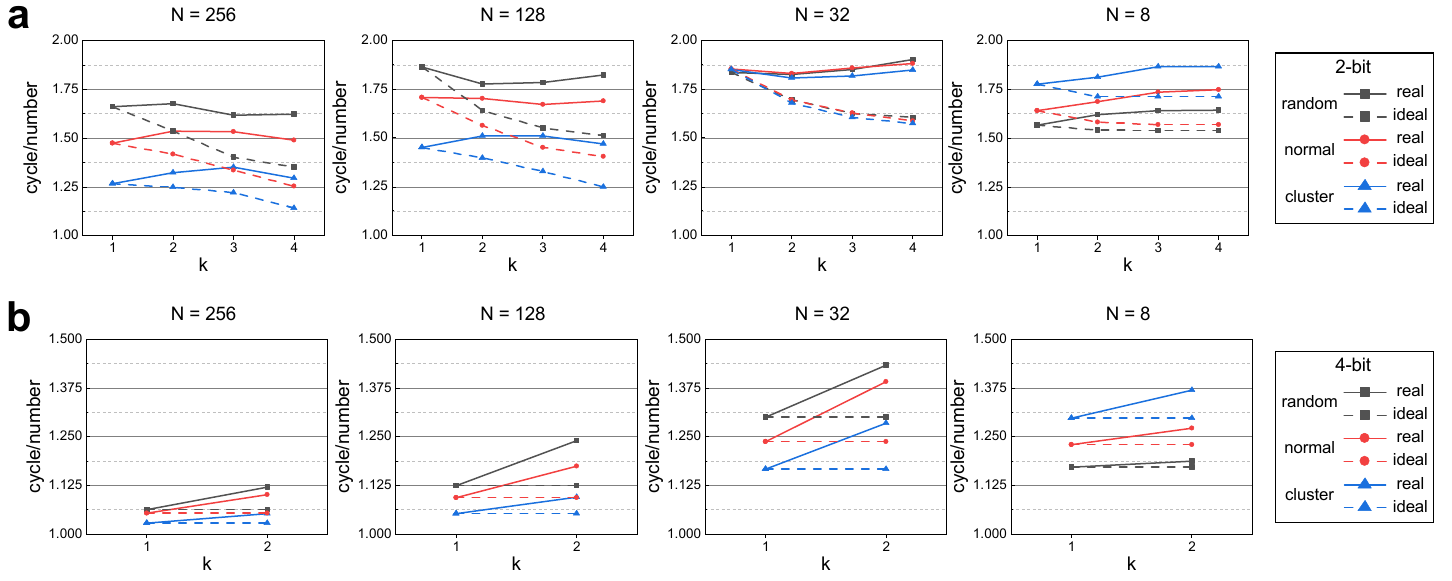}
    \caption{\textbf{Sorting speed for different data quantity using 8-bit unsigned numbers under 3 datasets in ideal and actual scenarios.}. \textbf{a,} ML-$2$-bit. \textbf{b,} ML-$4$-bit.}
    \label{fig:8b_tradeoff}
\end{figure}

\begin{figure}[hbt!]
    \centering
    \includegraphics[width = 1\linewidth]{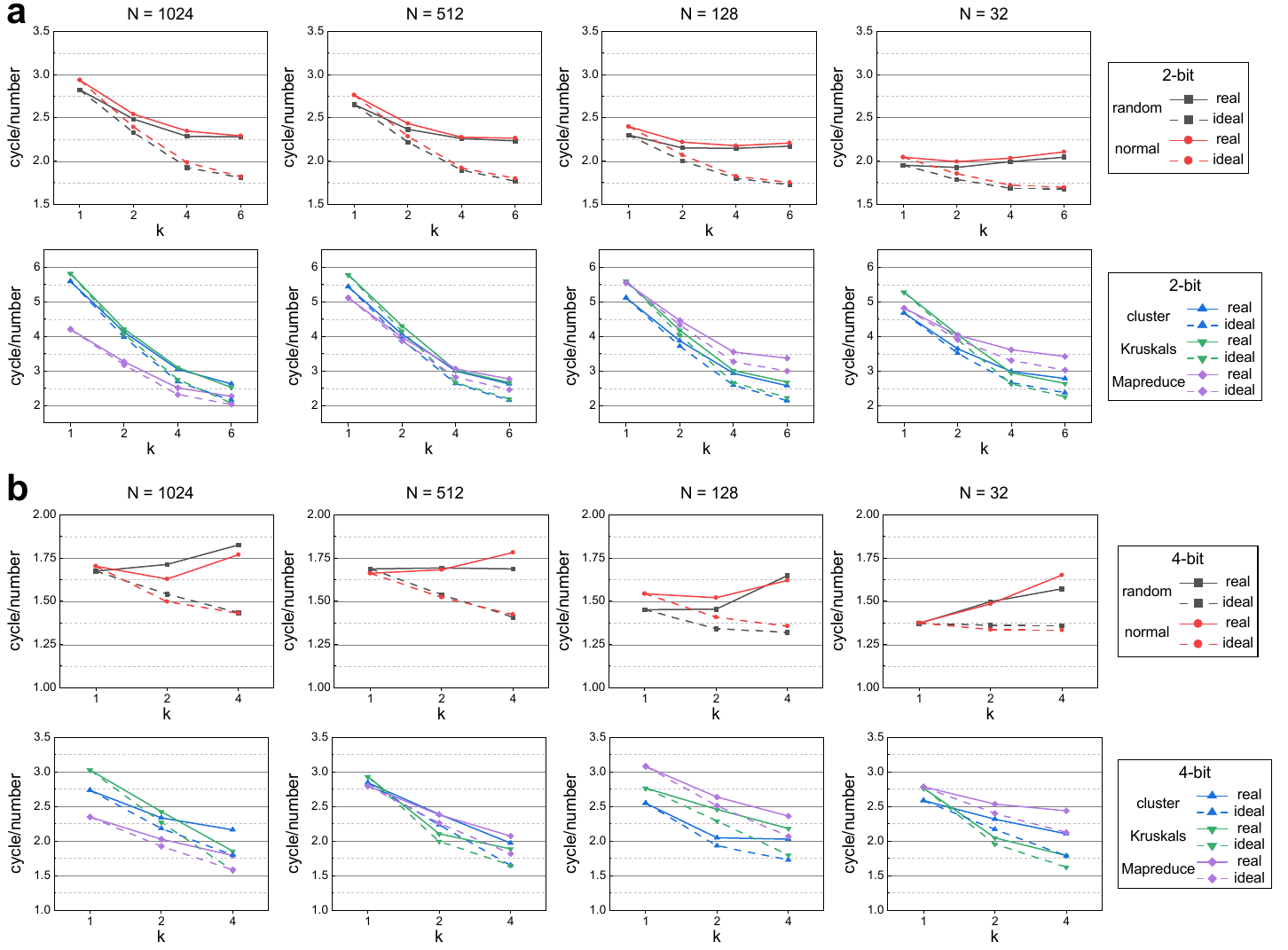}
    \caption{\textbf{Sorting speed for different data quantity using 32-bit unsigned numbers under 5 datasets in ideal and actual scenarios.}. \textbf{a,} ML-$2$-bit. \textbf{b,} ML-$4$-bit.}
    \label{fig:32b_tradeoff}
\end{figure}

To solve the redundant cycle problem, we can design a more strict state check mechanism for LIFOs, but it will cost more area and energy. We study the speedup differences between ideal and actual scenarios for different data quantity and data precision. The ideal scenario is that we use additional logic for more strict state check at LIFOs so that there is no redundant cycles to eliminate duplicate states, while the actual scenario is we do not use additional logic to eliminate duplicate states.

For 8-bit low precision datasets, we test different data quantity with three datasets, random, normal and clustered (\figurename~\ref{fig:8b_tradeoff}). When using ML-$2$-bit TNS sorting, we pick $k=1, 2, 3, 4$ for experiments because the DR tree has a depth of 4 (\figurename~\ref{fig:8b_tradeoff}a). From the experimental results, the sorting speed in actual scenario may not be faster when $k$ increases; but in ideal scenarios, the sorting speed goes up monotonically with $k$. This confirms our explanations before. Meanwhile, as $k$ increases, the differences between the two scenarios, representing the number of redundant cycles, are increasing gradually. For ML-$4$-bit TNS sorting (\figurename~\ref{fig:8b_tradeoff}b), we can observe similar behaviors. Note that the ML TNS sorting speed for 8-bit low precision cases reaches maximum at $k=1$ and we don't need a higher $k$ for faster sorting speed. 

For 32-bit high precision data, we experiment with ML-$2$-bit and ML-$4$-bit cases (\figurename~\ref{fig:32b_tradeoff}). For ML-$2$-bit, as $k$ increases, the speedup increase brought by $k$ is more significant than the speedup decrease caused by redundant cycles, although the number of redundant cycles actually increases with increasing $k$. When using ML-$4$-bit, the sorting speedup slows down as $k$ increases in random and normal datasets, but accelerates in other three datasets. In summary, there are more redundant cycles with $k$ increases; however, multi-level strategy can achieve approximately maximum speedup performance across different datasets at a lower $k$ so that there is no need to design extra circuitry to solve this redundant cycle problem.

\newpage
\section{Dijstra's Algorithm using TNS}\label{Details_Dijstra}

Dijkstra's algorithm is a widely used to find the shortest path between two nodes in a weighted graph. It has been adopted for subway route planning to find the shortest path between two subway stations. The pseudo-code for Dijkstra's algorithm is given in Algorithm~\ref{alg:dijkstra} below:

\begin{algorithm}[hbt!]
\caption{Dijkstra's algorithm for shortest path searching} 
\vspace{0.2in}
\hspace*{0.02in} {\bf Input:} 
distance map $d_{i, j}$, $i,j=1\rightarrow n$, start node $s$, end node $e$\\
\hspace*{0.02in} {\bf Output:} 
shortest distance map $D_{i}$, previous node map $C_{i}$, $i=1 \rightarrow n$
\begin{algorithmic}[1]

\For{$i=1 \rightarrow n$}
\State $D_{i}=+\infty$, add $i$ to $Q$
\EndFor
\State $D_{s}=0$

\While{$Q$ is not empty} 
\State remove $u$ from Q ($D_{u}$ is the minimum for index in $Q$)
\For{each neighbor $v$ of $u$ still in $Q$}
\State $S=D_{u}+d_{u,v}$
\If{$S<D_{v}$}
\State $D_{v}=S, C_{v}=u$
\EndIf
\EndFor
\EndWhile

\State \Return $D_{e},C$

\end{algorithmic}
\label{alg:dijkstra}
\end{algorithm}

\begin{figure}[hbt!]
    \centering
    \includegraphics[width = 1\linewidth]{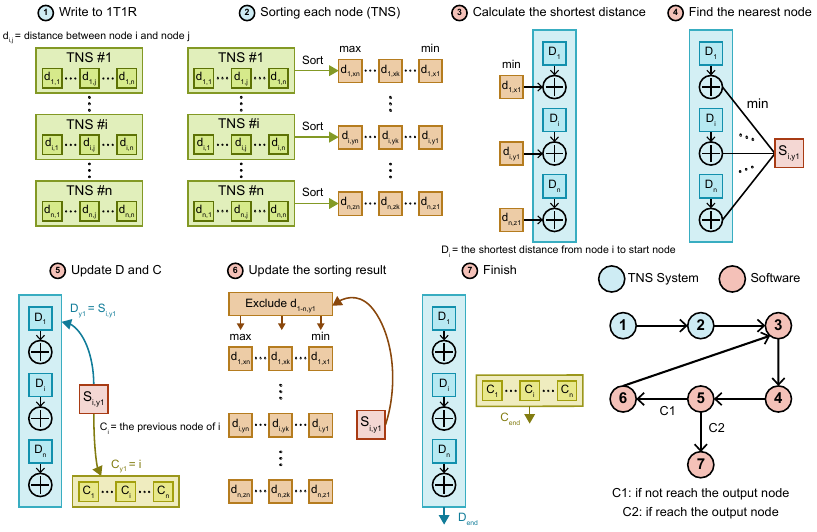}
    \caption{\textbf{Flow chart of the Dijstra's algorithm based on TNS.}}
    \label{fig:dijstra_flow}
\end{figure}

We map Dijstra's algorithm to TNS-based SIM system in \ref{subsec3.1} to achieve the shortest path search for 16 subway station nodes in Beijing. The entire task is accomplished by our memristor-based hardware and software co-designed SIM system. The details are shown in \figurename~\ref{fig:dijstra_flow}. Step 1 and 2 are implemented in memristor-based SIM hardware and other steps are finished in PC software. In step 1, we write the distances between station nodes and their neighbors into our 1T1R chip (\figurename~\ref{fig:dijstra}d). Then, we perform TNS sorting on each node separately and send the sorting results to PC software in Step 2. In step 3, for each station node, we find its nearest neighbor and add this distance to $D$. $D_{i}$ represents the shortest distance from node station $i$ to the start station node: $D$ of the start station node is 0 and the initial values of other station nodes are $+\infty$. In Step 4, we find the min value ($S_{i,y1}$), which is the closest distance to the start station node among all the values being summed in Step 3. 

In Step 5, we update $D$'s and $C$'s according to $S_{i,y1}$ ($D_{y1}=S_{i,y1}$, $C_{y1}=i$). At this point, we need to check whether the station node $i$ is the end station node. If not, we go to Step 6. According to the second subscript $y1$ of $S_{i,y1}$, we exclude the number with the same subscript in the sorting results ($d_{1 \sim n,y1}$) and then go back to Step 3, finding a new min value and starting the process again until reaching the end station node at Step 5. If we reach the end station node at Step 5, we output $D_{end}$ and $C_{end}$ to get the shortest distance and the corresponding path between start station node and end station node.

We give an example to illustrate the details of TNS-based Dijstra's algorithm, picking station node 3 (Xizhi Men) as the start station node and station node 13 (Jianguo Men) as the end station node (\figurename~\ref{fig:dijstra}a). We initialize the sorting configurations through PC and FPGAs and send the sorting results from our memristor-based hardware to PC software. The output $D_{13}$ shows that the shortest distance is $9.756km$ from Xizhi Men to Jianguo Men. Starting from the end station node, we find the previous station nodes step by step to obtain $C$. The path corresponding to the shortest distance is $node3\rightarrow node5\rightarrow node6\rightarrow node7\rightarrow node12\rightarrow node13$.

\newpage
\section{Details of In-Situ Pruning and Inference}\label{Detals_Pruning}

\begin{algorithm}[hbt!]
\caption{In-Situ Pruning Integrated with CIM-based MVM} 
\vspace{0.2in}
\hspace*{0.02in} {\bf Input:} 
input feature map $I$, weight $W$, number of weight $N$, pruning rate $p$\\
\hspace*{0.02in} {\bf Output:} 
output feature map $O$
\begin{algorithmic}[1]
\State $n=0$

\While{$n<=N\times p$} 
\State $I[\bf TNS_{min}$ $(abs(W))]=0$
\State $n=n+1$
\EndWhile

\State $O=$ $\bf CIM\_MVM$ $(I, W)$
\State \Return $O$

\end{algorithmic}
\label{alg:prune}
\end{algorithm}

In neural network inference, model weights with larger magnitudes generally have greater impacts on final results. Run-time tunable sparsity based on weight magnitudes are developed to skip unimportant weights\cite{raihan2020sparse} using one trained model for different applications. Here we use TNS to search for min values of weights magnitudes for in-situ pruning in memristor arrays. We set a tunable pruning rate $p$ to study the trade-off between network accuracy and energy consumption. Once we in-situ prune $N\times p$ ($N$ represents the total number of weights in a layer) weights with smallest magnitudes, we can integrate the process with subsequent CIM-based MVM inference as depicted in Algorithm~\ref{alg:prune}. .

\newpage
\section{PointNet++ with In-Situ Pruning and Inference}\label{Details_PointNet}

\begin{figure}[hbt!]
    \centering
    \includegraphics[width = 0.8\linewidth]{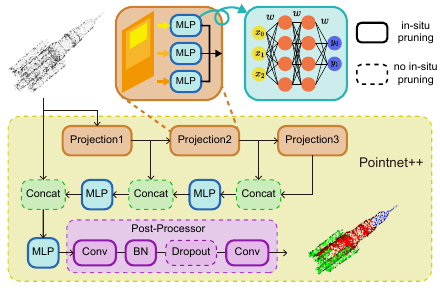}
    \caption{\textbf{Overall architecture of PointNet++. The parts where in-situ pruning being used are labeled in solid line boxes.}}
    \label{fig:pointnet arch}
\end{figure}

PointNet++ is a widely-used neural network for analyzing point cloud data \cite{qi2017pointnet++}. It introduces multi-scale processing and local region feature extraction to improve point cloud modeling efficiency. The PointNet++ network consists of 3 parts: Sampling, Grouping, and the Post Processor (\figurename~\ref{fig:pointnet arch}). Sampling contains three projections, each of which is used to extract information at different scales. Grouping consists of three concat and MLP layers for the aggregation of information. The Post Processor consists of two convolution and one batch normalization layers. 

\begin{table}[]
    \centering
    \includegraphics[width=0.8\textwidth]{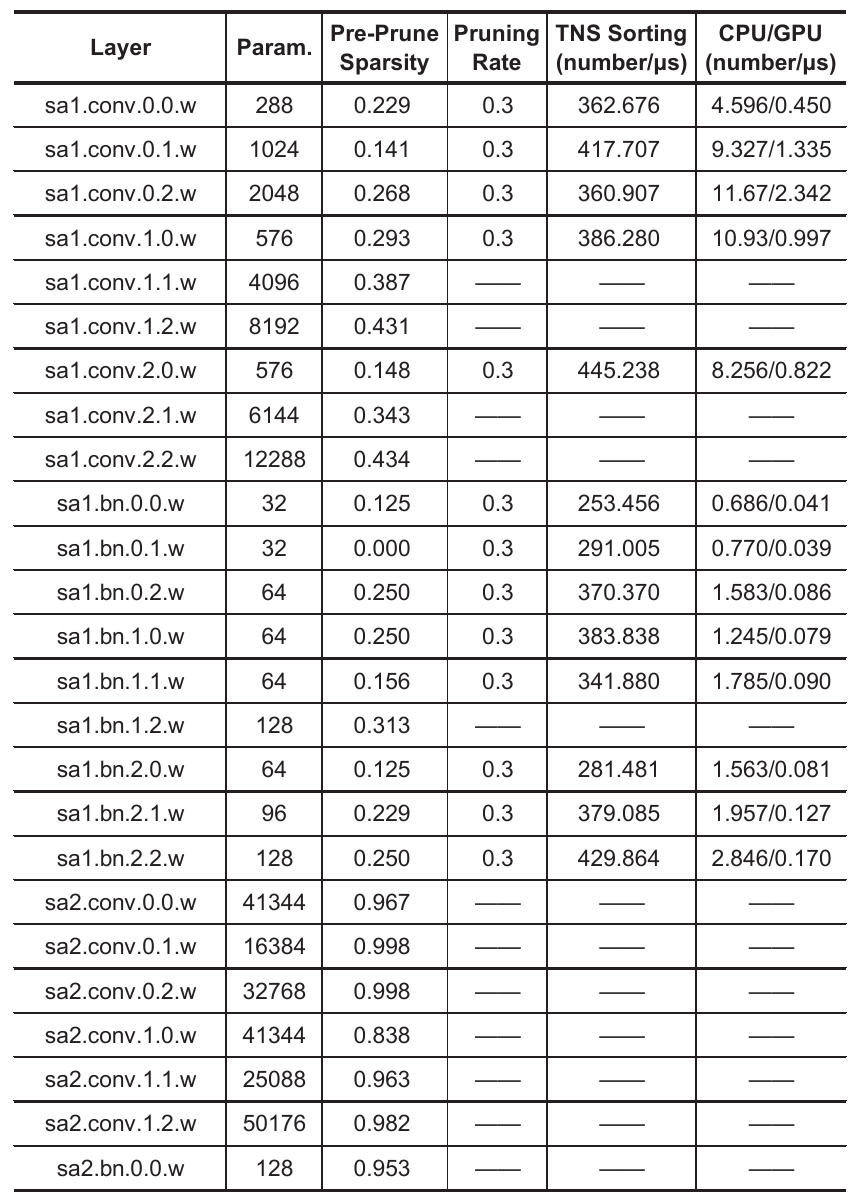}
    \caption{\textbf{Weight parameters, sparsity, pruning rates and sorting speed for all layers (Part 1) in PointNet++. We choose basic TNS with $k=2$ for TNS sorting.}}
    \label{tab:weight1}
\end{table}

\begin{table}[]
    \centering
    \includegraphics[width=0.8\textwidth]{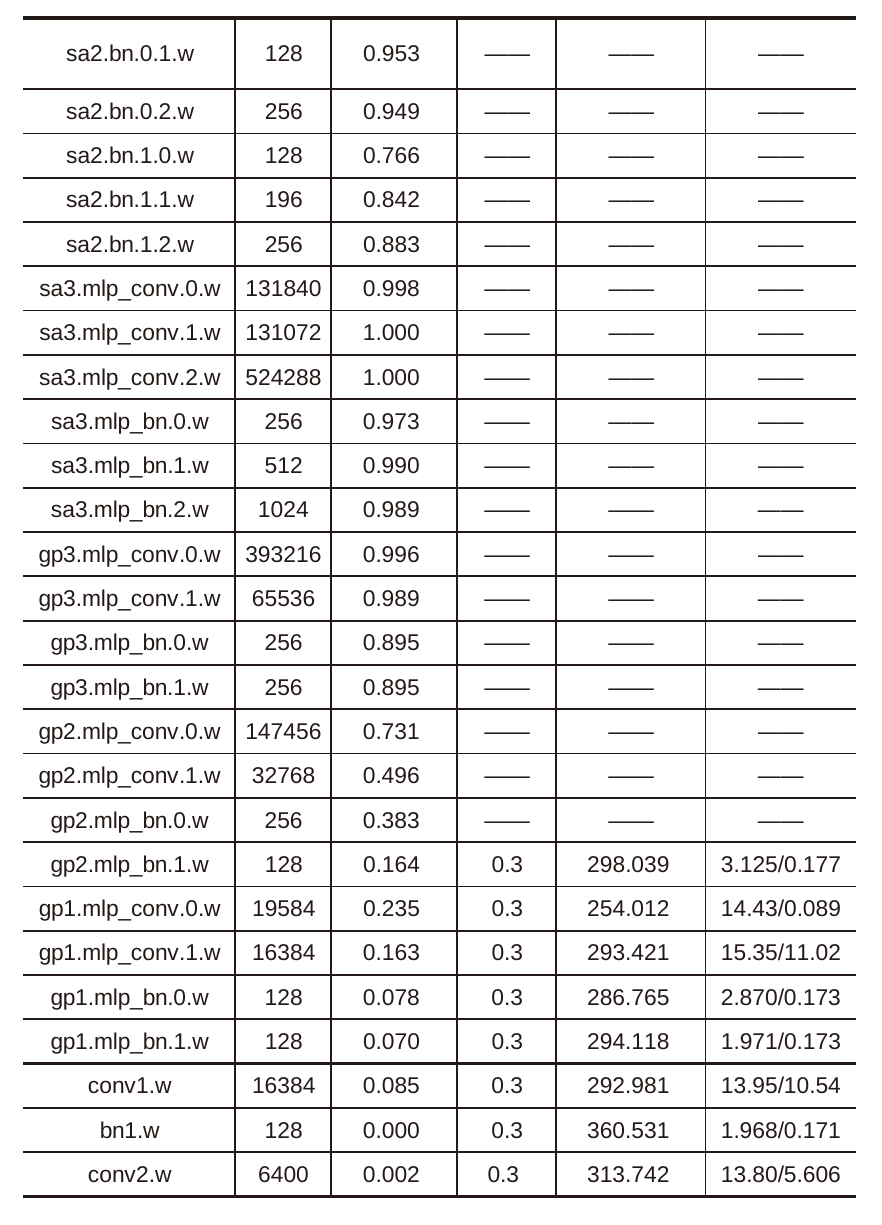}
    \caption{\textbf{Weight parameters, sparsity, pruning rates and sorting speed for all layers (Part 2) in PointNet++. We choose basic TNS with $k=2$ for TNS sorting.}}
    \label{tab:weight2}
\end{table}

\begin{figure}[hbt!]
    \centering
    \includegraphics[width = 1\linewidth]{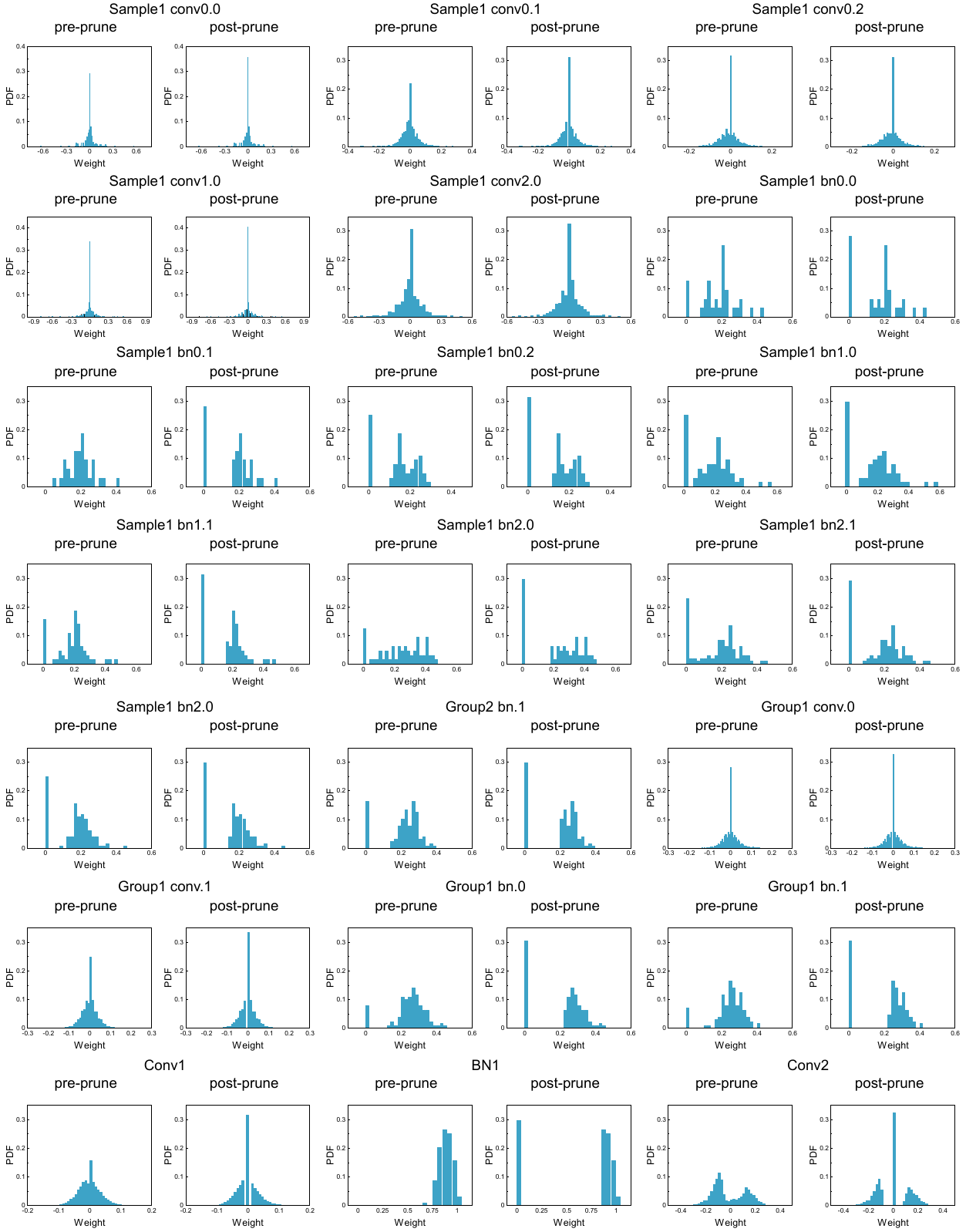}
    \caption{\textbf{Pre-pruned and post-pruned weight distributions of all pruning layers in PointNet++.}}
    \label{fig:distribution}
\end{figure}

We demonstrate the sparsity of weights in all above layers and prune weights with low magnitudes. The results are show in \tablename~\ref{tab:weight1} and \tablename~\ref{tab:weight2}. We quantify the weights into 8-bit sign-and-magnitude numbers and store them in our 1T1R array chip using the basic TNS algorithm ($k=2$). When the total number of minimums found by TNS reaches the pruning rate times the total number of weights in a layer, TNS stops working and all previously found minimums are discarded in subsequent MVM CIM calculations. Since conventional inference with run-time tunable sparsity are usually performed on CPUs/GPUs, we also measure the CPU/GPU latency for the same task for comparison purpose. As we can see from the sorting speed measurements (\tablename~\ref{tab:weight1} and \tablename~\ref{tab:weight2}), TNS achieves higher performance than the CPU/GPU across all layers. Note that the number of cycles needed for TNS sorting decreases as $N$ increases, because a large enough $N$ is able to cover all branching tree nodes on DR tree with a large number of repeated numbers that can be handled efficiently with our repeated number check module (\figurename~\ref{fig:Fig3}a). On the other hand, the clock frequency slows down with $N$ increases, but the sorting speedup still goes up since the DR latency reduction is more significant than frequency degradation.

\figurename~\ref{fig:distribution} shows the weight distribution of all pruning layers before and after pruning. Note that the weights of most layers follow a normal distribution, which shows consistently good sorting speedup performance as our previous evaluations (Supplementary Section~\ref{Details_Performance}).

\newpage
\section{PointNet++ with Multi-Level Devices}\label{ML_Accuracy}

\begin{figure}[hbt!]
    \centering
    \includegraphics[width = 0.6\linewidth]{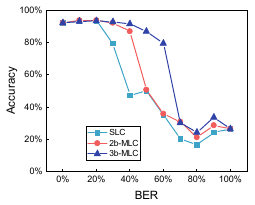}
    \caption{\textbf{The relationship between PointNet++ accuracy and BER under different hybrid precision conditions.}}
    \label{fig:ber}
\end{figure}

Failures to program memristor devices often lead to bit errors and affect the performance of in-situ pruning and inference for neural network. Bit errors in weights may also lead to incorrect sorting results for in-situ pruning. Here we measure the effects of bit error rates (BER) on PointNet++ with three different multi-level choices (\figurename~\ref{fig:ber}). PointNet++ exhibits high robustness with respect to BER and the network's inference accuracy only drops significantly when BER exceeds nearly $20\%$. The results show that in-situ pruning with multi-level devices has great potential for neural networks that has high robustness with respect to BER.

\newpage
\section{Experimental Comparisons}\label{Performance_Comp}

\begin{table}[hbt!]
    \centering
    \includegraphics[width=\textwidth]{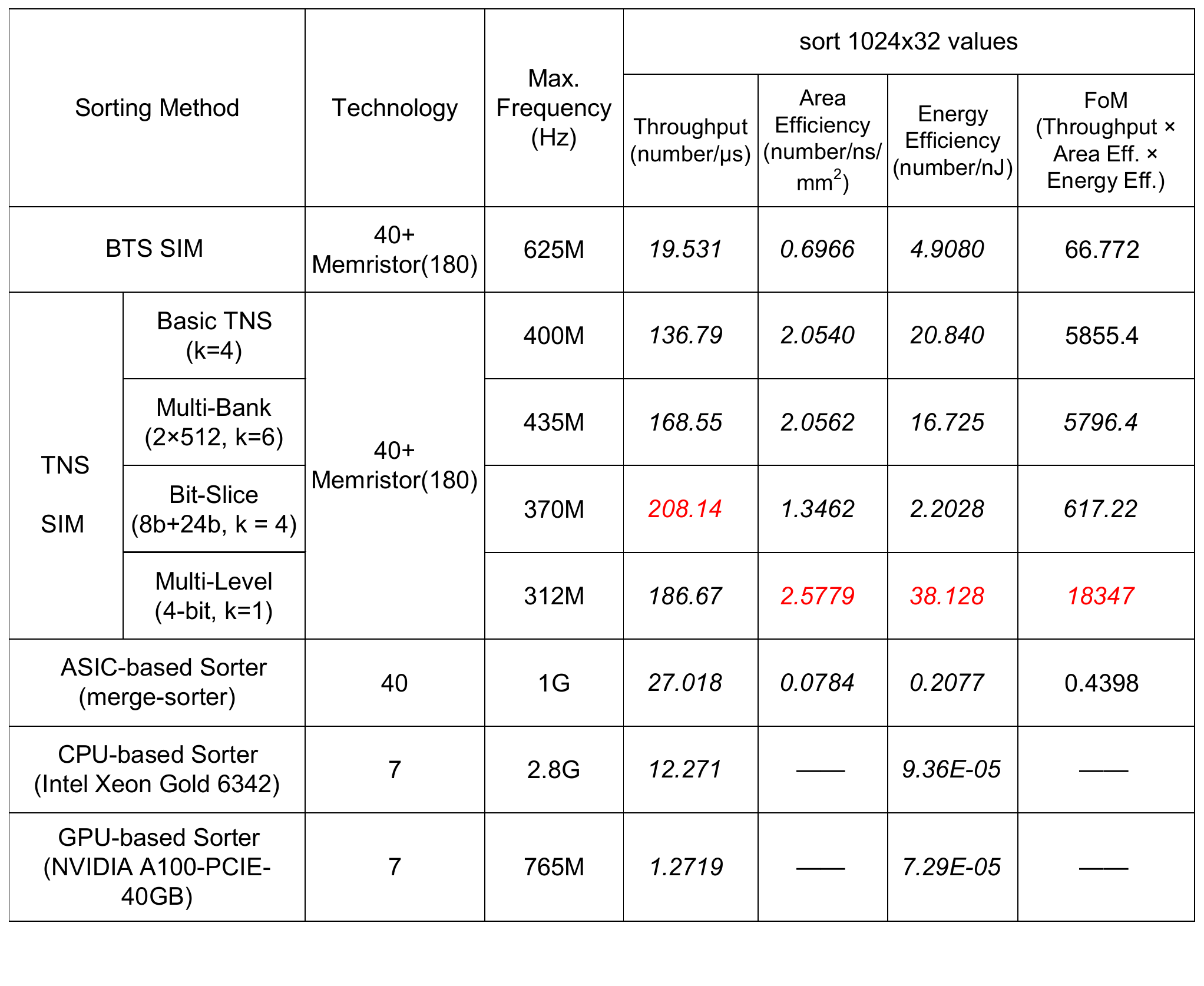}
    \caption{\textbf{Comparison based on experimental measurements between our memristor-based TNS/CA-TNS SIM system and ASIC-based, GPU-based, CPU-based sorting systems.}}
    \label{tab:finalcomp}
\end{table}

Taking sorting 1024 32-bit unsigned numbers as an example, we measure the metrics of different sorting systems for comparison. We measure the highest achievable operating frequencies and corresponding sorting throughput, area cost and energy consumption using our memristor-based TNS/CA-TNS SIM system. We also measure the performance of ASIC-based sorter (merge-sorter) as well as CPUs/GPUs-based sorters using the same dataset. The measurement results are shown in \tablename~\ref{tab:finalcomp}. We define a figure of merit (FoM) as throughput $\times$ area efficiency $\times$ energy efficiency and higher FoM stands for better sorting performance. 

Experimenting through different sorting configurations (with different paremeter $k$, multi-bank choices, bit-slice choices, and multi-level choices), we find that CA-TNS with BS strategy achieves the highest throughput of 208.14 number/$\mu s$ running at a maximum frequency of 370MHz. On the other hand, CA-TNS with ML strategy has the highest area efficiency of 2.5779 number/ns/mm$^2$ and energy efficiency of 39.128 number/nJ when running at a maximum clock frequency of 312MHz, reaching the best FoM of 18347. Compared to ASIC-based, CPU-based and GPU-based sorters using more advanced process technology, our memristor-based SIM system presents up to $3.32\times \sim 7.70 \times$ speedup, $6.23 \times \sim 183.5 \times$ energy efficiency and $2.23 \times \sim 7.43 \times$ smaller hardware area.

\end{document}